\documentclass[english,prc,fleqn, a4paper, twocolumn, superscriptaddress, showpacs]{revtex4}
\usepackage[T1]{fontenc}
\usepackage{amsmath}
\usepackage{graphicx}
\usepackage{amssymb}

\makeatletter

\newcommand{\noun}[1]{\textsc{#1}}
\providecommand{\tabularnewline}{\\}

\usepackage{graphicx}

\usepackage[squaren]{SIunits}
\usepackage{color}
\usepackage{array}
\usepackage{amssymb}
\usepackage{dcolumn}
\usepackage{psfrag}

\newcommand{\gr}[1]{\textcolor[gray]{0.6}{#1}}

\usepackage{babel}
\makeatother

\begin{document}

\title{Enhancement of the Deuteron-Fusion Reactions in Metals and its Experimental
Implications}

\author{A. Huke}

\email{huke@physik.tu-berlin.de, Armin.Huke@web.de}

\affiliation{Institut f\"{u}r Optik und Atomare Physik, Technische Universit\"{a}t
Berlin\\
Hardenbergstra\ss e 36, 10623 Berlin, Germany}

\author{K. Czerski}

\affiliation{Institute of Physics, University of Szczecin, Szczecin, Poland}

\affiliation{Institut f\"{u}r Optik und Atomare Physik, Technische Universit\"{a}t
Berlin\\
Hardenbergstra\ss e 36, 10623 Berlin, Germany}

\author{P. Heide}

\affiliation{Institut f\"{u}r Optik und Atomare Physik, Technische Universit\"{a}t
Berlin\\
Hardenbergstra\ss e 36, 10623 Berlin, Germany}

\author{G. Ruprecht}

\affiliation{TRIUMF, Vancouver, B.C., Canada}

\affiliation{Institut f\"{u}r Optik und Atomare Physik, Technische Universit\"{a}t
Berlin\\
Hardenbergstra\ss e 36, 10623 Berlin, Germany}

\author{N. Targosz}

\affiliation{Institute of Physics, University of Szczecin, Szczecin, Poland}

\author{W. \.{Z}ebrowski}

\affiliation{Institute of Physics, University of Szczecin, Szczecin, Poland}

\begin{abstract}
Recent measurements of the reaction $^{2}$H(d,p)$^{3}$H in metallic
environments at very low energies performed by different experimental
groups point to an enhanced electron screening effect. However, the
resulting screening energies differ strongly for divers host metals
and different experiments. Here, we present new experimental results
and investigations of interfering processes in the irradiated targets.
These measurements inside metals set special challenges and pitfalls
which make them and the data analysis particularly error-prone. There
are multi-parameter collateral effects which are crucial for the correct
interpretation of the observed experimental yields. They mainly originate
from target surface contaminations due to residual gases in the vacuum
as well as from inhomogeneities and instabilities in the deuteron
density distribution in the targets. In order to address these problems
an improved differential analysis method beyond the standard procedures
has been implemented. Profound scrutiny of the other experiments demonstrates
that the observed unusual changes in the reaction yields are mainly
due to deuteron density dynamics simulating the alleged screening
energy values. The experimental results are compared with different
theoretical models of the electron screening in metals. The Debye-H\"{u}ckel
model that has been previously proposed to explain the influence of
the electron screening on both nuclear reactions and radioactive decays
could be clearly excluded.
\end{abstract}


\maketitle

\section{INTRODUCTION}

The cross section for nuclear reactions between charged particles
at low energies is mainly determined by the penetration probability
through the Coulomb barrier, which results in a steep exponential
decrease towards lower energies. At sufficiently low energies, however,
this decrease is slowed down due to screening the Coulomb barrier
by the inevitable presence of surrounding electrons. The electron
screening was originally taken into account for nuclear reactions
preceding in dense astrophysical plasmas in the interior of stars
\citep{salpeter54} where the nuclear reaction rates can be increased
even by many orders of magnitude. For laboratory investigations of
nuclear reactions at very low energies, this effect was theoretically
described \citep{assenbaum87} and experimentally observed in different
fusion reactions on gas targets, e.g.~\citep{rolfs95}. The corresponding
enhancement of the nuclear cross section could be explained by the
gain of electron binding energies between the initial distant atoms
and the final fused atom. This was attributed to the raise of the
kinetic energy of colliding nuclei and called electron screening energy.
For the first time, the electron screening effect resulting from much
more important for astrophysical applications free electrons was investigated
in the d+d fusion reactions taking place in metallic environments
\citep{volos98,europhys01,dis}. The experimentally determined screening
energies for some heavier metals were one order of magnitude larger
than the gas target value \citep{greife95} and larger by a factor
of about four than the theoretical predictions \citep{ichimaru93}.
These experimental results were also confirmed by other authors \citep{yuki98,kasagi02,rolfs02,rolfs02b,rolfs03,rolfs04}. 

Meanwhile, the electron screening effect in the d+d fusion reactions
has been studied for over 50 different metals and some insulators
\citep{rolfs02b,rolfs03,rolfs04} allowing, in principle, for a systematic
study of the target material dependence of the electron screening
energy. Unfortunately, there are some discrepancies between experimental
values obtained by different groups \citep{NPAII06b}. They probably
arise from some experimental systematic uncertainties connected with
oxidation of the target surface or with a high mobility of the implanted
deuterons under beam irradiation, which can lead to unstable deuteron
density profiles within the target. Both effects play a crucial role
for the experimental determination of the screening energies \citep{dis,nimb06}.
The basic quantity received from the experiment is the nuclear reaction
yield which is given for a thick target as an integral over the range
of the projectiles $Y=\int_{0}^{R}\left[n\sigma\right]dx$ with the
target nuclei density distribution $n$ and the cross section $\sigma$.
So deviations in the observed yield have the two principal causes:
changes in the deuteron density profile and modification of the cross
section, probably by the screening effect, which are merged in the
integrant product $\left[n\sigma\right]$. Some standard experimental
difficulties have been already discussed in our preceding paper \citep{nimb06}
where an especially adapted data acquisition and analysis method,
allowing us to discern between changes in $n$ and $\sigma$, has
been presented in a systematic manner. Based on this method, we report
here some new experimental results and estimate experimental uncertainties
of previous experiments. We additionally compare data obtained by
different groups and discuss systematic errors of applied experimental
and analytical methods.

From the theoretical point of view, the large number of experimental
data corrected for the discussed experimental uncertainties enables
a comparison with theoretical predictions. The first ab-initio quantum
mechanical calculation of the screening energy in a crystal environment
has been recently performed using realistic wave functions \citep{epja08}.
However, the results are still unsatisfying because of the very high
demand for computational power limiting the model accuracy. Thus,
the self-consistent dielectric function theory developed previously
\citep{europhys04} will be used here for the calculation of the screening
energy contributions coming not only from free electrons but also
from bound electrons of reacting nuclei and host metals. Additionally,
the interaction with the crystal lattice will be included. The theoretical
results will be extended for comparison with the last experimental
studies of the electron screening in nuclear reactions between heavier
nuclei \citep{kasagi04,rolfs05c,rolfs06b} and in radioactive decays
\citep{limata06,rolfs06c,ruprecht06a}. On the other hand, it has
recently been suggested that the enhanced electron screening can be
explained within the classical Debye-H\"{u}ckel model \citep{rolfs03}.
The idea has been supported by an observation of the predicted inverse
proportionality of the experimental screening energies to the square
root of the absolute target temperature \citep{rolfs05b,rolfs06}.
As a consequence one could expect a magnification of the $\alpha$
and $\beta^{+}$ decay rates when radioactive sources would be put
in metals at cryogenic temperatures. Even though the Debye screening
cannot be applied to strongly coupled electron plasmas, as metals
at moderate temperatures are, the suggestion has found much public
interest \citep{rolfs_public2,rolfs_public3,rolfs08,rolfs06c}. Thus,
both experimental and theoretical aspects of the temperature effect
of the electron screening will be subject of a critical discussion
clearly showing the inapplicability of the Debye-H\"{u}ckel model
for these issues.

\section{\label{sec:setup}Experimental set-up, data acquisition and analysis}

\begin{figure*}
\begin{centering}
\includegraphics[width=1\textwidth]{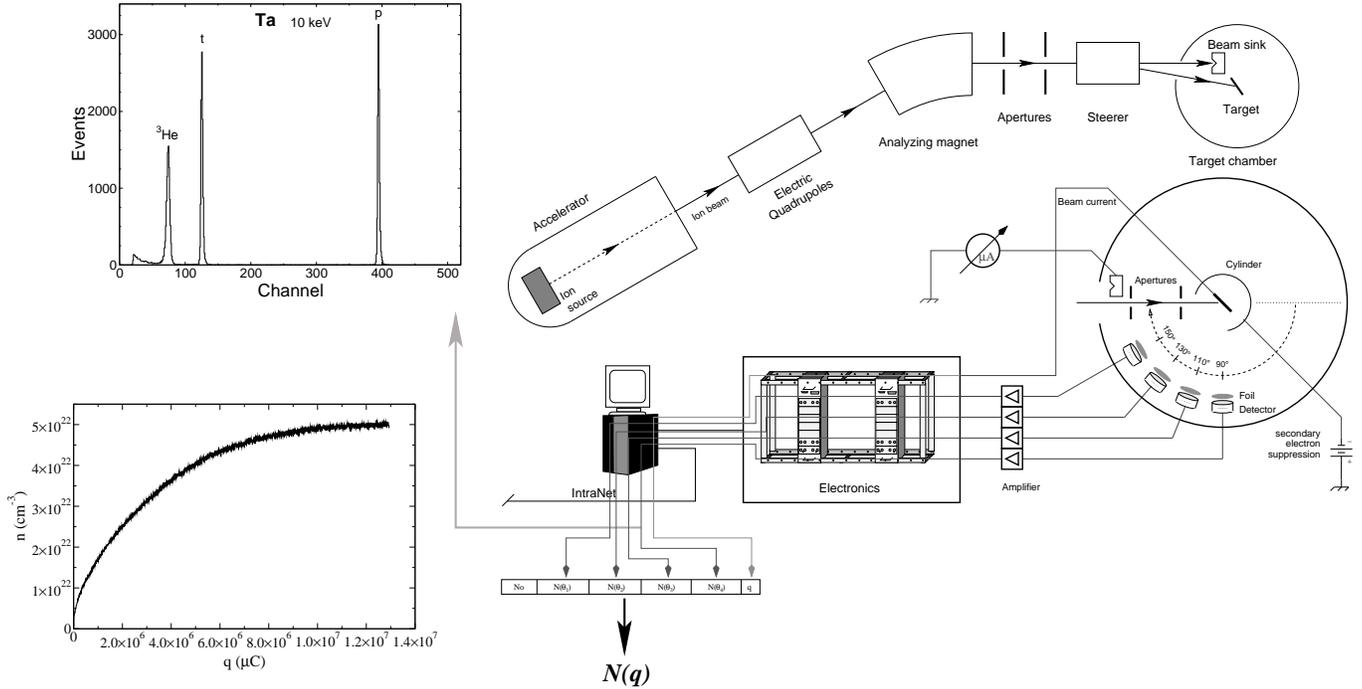}
\par\end{centering}

\caption{\label{fig:experiment}Experimental set-up}

\end{figure*}

The experiments have been carried out at an accelerator optimized
for low energy beams. Fig.~\ref{fig:experiment} illustrates the
principal set-up and the data acquisition system. %
 The accelerator consists of a radio frequency ion source, an acceleration
line powered by a highly stabilized $60\,\kilo\volt$ supply and subsequent
electric quadrupoles for focusing and a magnetic dipole for beam analysis.
The beam impinges onto a Faraday cup just inside the target chamber
where beam adjustment can be done without disturbing the deuteron
density in the targets. A horizontal magnetic steerer is used to deflect
the beam onto the target, such removing neutral particles and contaminations
carried along by the beam. A cylinder box set to a negative potential
surrounds the target in order to suppress secondary electrons. The
isolated target holder is connected to a current integrator. The targets
were disks made from different pure metals becoming self-implanted
deuterium targets under the beam irradiation. Four Si-detectors at
the laboratory angles of 90$^{\circ}$, 110$^{\circ}$, 130$^{\circ}$
and 150$^{\circ}$ were used for the detection of all charged particles,
p, t, $^{3}$He, of the reactions $^{2}$H(d,p)t and $^{2}$H(d,n)$^{3}$He.
The detectors needed to be shielded from the backscattered deuterons
in order to prevent a congestion of them and the data acquisition
system. Therefore grounded Al-foils of thicknesses from $120-150\,\micro\gram\per\centi\meter\squared$
were placed in front of the detectors. The thickness is sufficient
to block deuterons up to $60\,\kilo\electronvolt$ while all other
ejectiles could pass. The detector voltage pulses travel through pre-amplifiers
and spectroscopic amplifiers. The signals are digitized by four ADCs
in an embedded VME system connected to a computer which automatically
integrates the proton lines of the spectra in fixed time intervals
\footnote{down to $10\,\second$ limited by the serial line%
} and records the four differential counting numbers $N\left(\theta\right)$
and the charge $q$ of the integrated beam current at the target in
a file which then can be further processed. An example spectrum is
shown in fig~\ref{fig:experiment}; all ejectile lines are clearly
identifiable. Due to the anisotropic angular distribution of the ejectiles
of the d+d fusion reactions even at the lowest energies, a total counting
number $N$ is calculated providing the tabulated function $N\left(q\right)$
which is the basic quantity for the further data analysis.

\begin{figure}
\begin{centering}
\includegraphics[width=1\columnwidth]{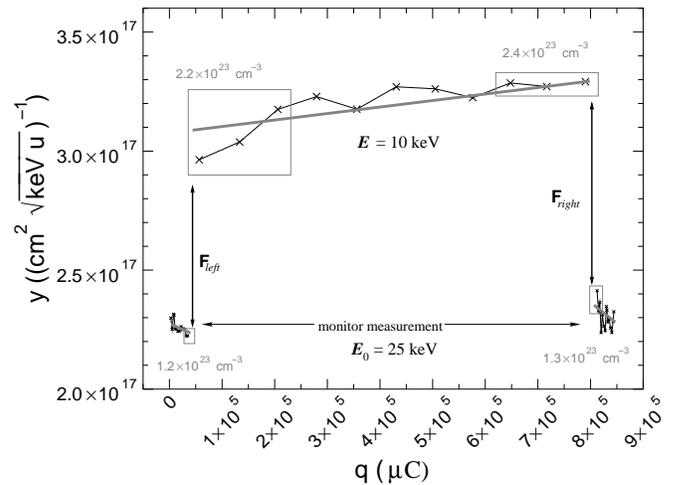}
\par\end{centering}

\caption{\label{fig:expprocedure}Analysis procedure at the example of Zirconium
at $10\,\kilo\electronvolt$}

\end{figure}

\begin{figure}
\begin{centering}
\includegraphics[width=1\columnwidth]{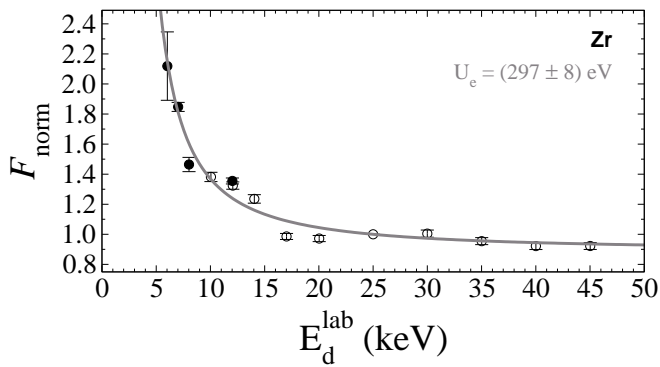}
\par\end{centering}

\caption{\label{fig:Fnx}Exemplary results for the enhancement factor $F_{\mathrm{norm}}$.
Screening enhancement for Zr theoretically described by the curve
with the single parameter $U_{e}$.}

\end{figure}

Correspondingly, the experimental reaction yield is given by \begin{equation}
Y\left(E\right)=\frac{ze}{\varepsilon}\frac{dN}{dq}\label{m:yieldexp}\end{equation}
 where the number of impacting projectiles is already substituted
by their charge, $\varepsilon$ is the detector efficiency and $z$
the charge state of the projectile. On the other hand the yield can
be calculated for an infinitely thick target (regarding the projectile
range $R$) by \begin{equation}
Y_{\mathrm{theo}}\left(E\right)=\int_{0}^{R}n\cdot\sigma\left(E\left(x\right)\right)dx\label{m:yieldtheo}\end{equation}
 with the number density of the target nuclei $n$ and the cross section
$\sigma$. Unlike other chemical compounds the small hydrogen atoms
are not trapped in firm chemical bonds with metals. The hydrogen density
is not bound to a fixed stoichiometric ratio and can and indeed does
change under ion irradiation. Changes in the yield may now originate
from both the deuteron density and the cross section and need to be
discriminated. The density is here a function of the target depth,
the projectile energy, the implanted charge, the beam flux and other
target material dependent and environmental conditions. The tabulated
function $N\left(q\right)$ provided by our data acquisition system
makes it possible to retain the differentiation in (\ref{m:yieldexp})
and thereby gain information on the charge development of a depth
averaged density $n\left(q\right)$. So assuming depth homogeneity
of the deuteron density in (\ref{m:yieldtheo}) the depth $x$ can
be substituted by the projectile energy $E$ with the stopping power
differential equation \citep{ziegler77} \begin{equation}
\frac{dE}{dx}=-\left(c_{M}+\frac{n(q)}{n_{D}}c_{D}\right)\sqrt{E}\label{m:abbdgl}\end{equation}
 where $c_{M}$ and $c_{D}$ are the stopping power coefficients in
the metal and in hydrogen, $n_{D}$ the appendant hydrogen density.
Applying this substitution one arrives at a motivation and an expression
for the \emph{reduced yield} \citep{europhys01,dis,nimb06} \begin{equation}
y(E;q):=\frac{Y(E;q)}{\int_{0}^{E}\frac{\sigma(E)}{\sqrt{E}}dE}=\gr{\frac{n(q)}{c_{M}+\frac{n(q)}{n_{D}}c_{D}}}\times F\left(E\right)\;.\label{m:redy}\end{equation}
 Since both the cross section in the metallic environment and the
deuteron density are unknown the yield need to be set in relation
to a known gas target cross section. We therefore chose the parameterization
from \citep{brown90} because it has the highest precision. It forms
together with the low energy function ($\sqrt{E}$) of the stopping
power (\ref{m:abbdgl}) the integral in the denominator on the right
hand side. The gray printed expression is per se a constant. So if
the reduced yield is not constant it is based on deviations of the
prescribed progression in the cross section or in the functional dependence
of the stopping powers or changes in the density. It is a sensitive
measure for such deviations but the distinction of the possible reasons
is a matter of reasonable interpretation. Fig.~\ref{fig:expprocedure}
shows plots of the reduced yield at two different energies. %
 One can see long term changes in the individual measurements indicated
by the straight lines. These are attributed to changes in the deuteron
density profiles scattered by the counting statistics, of course.
In contrast, the large discontinuities of the reduced yield at the
switching of the beam energy result from a modification of the cross
section. This is taken into account by the enhancement factor $F\left(E\right)$
in (\ref{m:redy}). Since the absolute quantity of the deuteron density
is unknown for the practical analysis a normalized enhancement factor
is defined \begin{equation}
F_{\mathrm{norm}}(E):=\frac{y(E)}{y(E_{0})}=\frac{F(E)}{F(E_{0})}\label{m:Fnorm}\end{equation}
 with the normalization energy $E_{0}$ which is chosen to be $25\,\kilo\electronvolt$
for the monitor measurements. The gray rectangles indicate the points
from which the experimental error for $F_{\mathrm{norm}}$ is inferred.
Thus not only errors from the counting statistics but also from long
term changes of the density are included. Results obtained for different
projectile energies are displayed in Fig.~\ref{fig:Fnx}. %
 Assuming electron screening as the reason for the increase of $F_{\mathrm{norm}}$
and adopting $U_{e}$ as a kinetic energy shift parameter called the
screening energy in the cross section \citep{assenbaum87} of the
yield one receives \citep{europhys01,dis,nimb06} \begin{equation}
F(E)=\frac{\int_{0}^{E}\frac{\sigma(E+2U_{e})}{\sqrt{E}}dE}{\int_{0}^{E}\frac{\sigma(E)}{\sqrt{E}}dE}\label{m:Fscr}\end{equation}
 for the screening enhancement factor of thick target yields %
\footnote{The screening energy $U_{e}$ should only be applied to the Coulomb
barrier penetration in $\sigma$, see \citep{nimb06,europhys04}.
The correction becomes only important for far lower beam energies.%
}. The factor 2 arises from the CM-Lab-transformation. So $F$ is an
enhancement factor for thick targets in analogy to the enhancement
factor for thin targets from \citep{assenbaum87} \begin{eqnarray}
f(E_{\mathrm{CM}}) & := & \frac{\sigma\left(E_{\mathrm{CM}}+U_{e}\right)}{\sigma\left(E_{\mathrm{CM}}\right)}\label{m:kontext.enhancedef-f}\\
 & = & \frac{\frac{1}{E_{\mathrm{CM}}+U_{e}}S\left(E_{\mathrm{CM}}+U_{e}\right)e^{-2\pi\eta\left(E_{\mathrm{CM}}+U_{e}\right)}}{\frac{1}{E_{\mathrm{CM}}}S(E_{\mathrm{CM}})e^{-2\pi\eta(E_{\mathrm{CM}})}}\nonumber \\
 & \simeq & e^{\left(\pi\eta\left(E_{\mathrm{CM}}\right)\frac{U_{e}}{E_{\mathrm{CM}}}\right)}\qquad,\, U_{e}\ll E_{\mathrm{CM}}\:,\nonumber \end{eqnarray}
 using the S-factor parametrization of the cross section with the
Sommerfeld parameter $\eta$ in the second line and applying an approximation
in the third line, which demonstrates its qualitative behaviour as
a roughly exponential increase for decreasing energies. The corresponding
curve in Fig.~\ref{fig:Fnx} obtained for a fitted value of $U_{e}$
supports the screening hypothesis. Our data analysis procedure is
thus independent of the absolute value of the deuteron densities inside
the targets and the stopping power coefficients which otherwise would
introduce errors of $10-20\%$. The functional dependency of the stopping
powers on the energy $\sqrt{E}$ has been repeatedly confirmed, see
\citep{moller04} and references therein. The reduced yield can be
used to calculate a deuteron density estimate by solving (\ref{m:redy})
for $n\left(q\right)$ and supposing $F=1$ \citep[Eq.\ (10)]{nimb06}.
Only for this purpose the stopping power coefficients are explicitly
required. A corresponding density plot for an initial implantation
in Al is shown in Fig.~\ref{fig:experiment}. The numbers above the
gray boxes in Fig.~\ref{fig:expprocedure} are density estimates
for that areas.

This is in brief the basic experimental procedure as of \citep{europhys01,dis,nimb06}.
For the study of the electron screening effect two experimental campaigns
were executed. Since the special physico-chemical properties of the
hydrogen compounds and the beam induced chemical reactions at the
target heavily influence the obtained results \citep{dis,nimb06},
the second more extensive campaign needed to investigate these interfering
effects \citep[Sec.\ 4]{nimb06} which are sketched in a concise survey
in the next section \ref{sec:expspecial}.

\section{\label{sec:expspecial}Experimental specialties and pitfalls}

The investigation of nuclear reaction cross sections on deuterium
in metals should be performed at the lowest possible energies. This
means that the composition of the topmost atomic layers of the metallic
target is of crucial importance because of the quickly decreasing
range of the beam ions, considerably below $1\,\micro\meter$. This
exactly is unusual for experimental nuclear physics. The usual set-ups
in experimental nuclear physics are constructed in high vacuum technology.
But here the contained water vapour from the surfaces of all materials
leads under ion impact to a progressing oxidation of the target metal
because of the stronger electron negativity of oxygen in comparison
to hydrogen. Hence, hydrogen is contained in metal oxides only in
segregation at low and unstable densities. Consequently, the oxidation
diminishes and eventually destroys the screening effect with the growing
thickness of the metal oxide layer. Carbon hydrides contained in HV
systems pose another problem leading to carbon layers on the target
as will be discussed below. In such a way generated alterations in
the depth profile of the deuteron density distribution in the target
is the singular dominating error source for the observed enhancement
and the inferred screening energies. Our vacuum system is made of
aluminium with elastomer gaskets pumped by turbo molecular pumps with
auxiliary oil lubricated two stage rotary vane pumps and LN$_{2}$
cooled cryogenic traps \citep[Fig.\ 1]{nimb06}. A residual gas analyzer
(RGA) was used in order to monitor the composition of the residual
gas in the vacuum. This is here merely a concise presentation; for
a more extensive description see \citep{nimb06}.

\begin{figure}
\begin{centering}
\includegraphics[width=1\columnwidth]{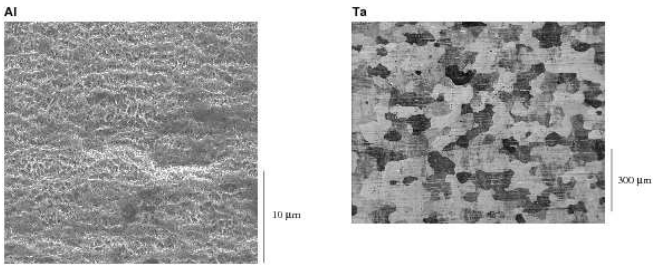}
\par\end{centering}

\caption{\label{fig:emic}Scanning electron microscopic pictures of target
surfaces. Left: Symptoms of embrittlement for Al. Right: Beginning
layer formation for Ta in island growth mode.}

\end{figure}

\begin{figure}
\begin{centering}
\includegraphics[width=1\columnwidth]{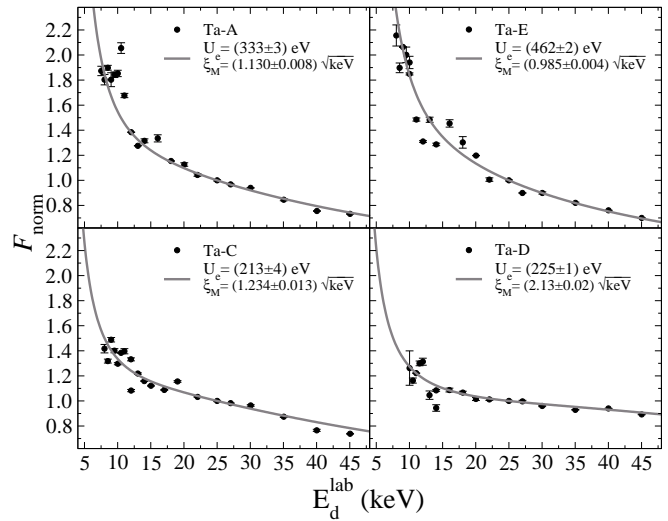}
\par\end{centering}

\caption{\label{fig:fituxim} Effects of different surface compositions on
the inferred screening energy for Ta. Ta-A has a small C-excess, Ta-E
has slight C-traces, Ta-C a thick C-layer, Ta-D a thick MO$_{x}$-layer.}

\end{figure}

In accordance to the literature about HV systems the main constituent
of the residual gas is water. Water vapour is due to its extraordinarily
high dipole moment very adhesive to solids and is hence chemisorbed
to surfaces. Now under the ion irradiation several processes are enabled.
Via heating and phonon excitation at the surface the beam provides
the activation energy for dissociative chemisorption of the water
molecule, i.e.~the protons are splitted off and the remaining oxygen
radical forms a chemical bond to the metal atoms. Essentially, the
same happens by direct impact excitation of the water molecule by
the ions. The hydrogen implantation into the metal causes aside from
the usual surface deterioration a in depth destruction of the crystal
integrity of the material known as embrittlement which always occurs
if the hydrogen loading rate is too high and not proceeding in thermal
equilibrium \citep{mhydrides68}. Thus, the surface is fractalized
and the oxidation can progress into the bulk of the metal quickly
creating a thick metal oxide layer. Fig.~\ref{fig:emic} contains
as an example for it a picture of the surface of an Al target which
turned into a sponge like structure. %
 The rate of the oxidation process depends on the concrete form of
the mutual interaction potential between the water molecule and the
surface atoms, establishing a material dependency. The energy supply
of the beam enables these processes even for the noble metals. Albeit
generally spoken, more reactive metals apt more to oxidation and embrittlement
while for the latter the structural difference between the metal and
the metal hydride is more important. Aside from the overall beam heating
the energy of the projectiles is also important because lower energy
projectiles are more effective at the surface \citep{ensinger97}.
The partial pressure of water in HV is so high that there are ample
supplies for the surface chemical reactions. The hit rate of water
molecules with a sticking coefficient of almost one is in comparable
orders of magnitude as usual beam currents of $10-100\,\micro\ampere$.
This implies a dependence on the ion flux, too. There are two counteracting
processes: Sputtering and thermal or ion stimulated desorption. The
sputtering yield of the lightweight deuterons is far too low in order
to keep the surface clean with the resulting sputtering rate. One
would expect that an increased temperature of the surface would increase
the desorption rate of the water molecules. If the activation energy
barrier for dissociative chemisorption of water is positive an increased
temperature yet proliferates the oxidation%
\footnote{see e.g.~\citep{zangwill88} or any surface physics textbook%
}. Similar is valid for ion stimulated desorption/chemisorption. Such
again depends on the interaction potential but usually oxidation prevails.
Unless UHV systems equipped for entire baking are used the oxidation
cannot be avoided. A deuteron irradiation of only $1\,\coulomb$ is
enough to produce a considerable metal oxide layer, see \citep[Fig.\ 6]{nimb06}.
There is, however, a process that is nonetheless able to prevent oxidation:
large carbon hydride molecules, e.g.~backstreaming from the forepumps,
can be physisorbed at the surface, cracked up and the carbon atoms
can react with the oxygen radicals to carbon monoxide keeping in that
way the surface clean. Differently from water, carbon hydrides are
physisorbed to surfaces. The strength of this weaker bond increases
with growing molecular mass. The ratio of absorption and desorption
under the ion irradiation has similar dependencies. An evidence for
this chemical surface reaction is the detection of a considerable
CO fraction by the RGA which was below the detection threshold without
beam irradiation \citep[Fig.\ 7]{nimb06}. These processes were thoroughly
explored by the regulated infusion of decane with monitoring feedback
as the main part of the second experimental campaign. The surface
can only be kept clean if the fraction of water and carbon hydrides
in the residual gas are in an equilibrium which is of course also
dependent on prementioned parameters. If the fraction of carbon hydrides
is too low the surface will oxidize. If it is too high a carbon layer
will build up. Both is essentially irreversible. Fig.~\ref{fig:fituxim}
shows some of the results of these experiments for Ta demonstrating
the high spread in the inferred screening energies depending on the
surface composition which were verified by electron dispersive X-ray
micro analysis (EDX). %

\begin{figure}
\begin{centering}
\includegraphics[width=1\columnwidth]{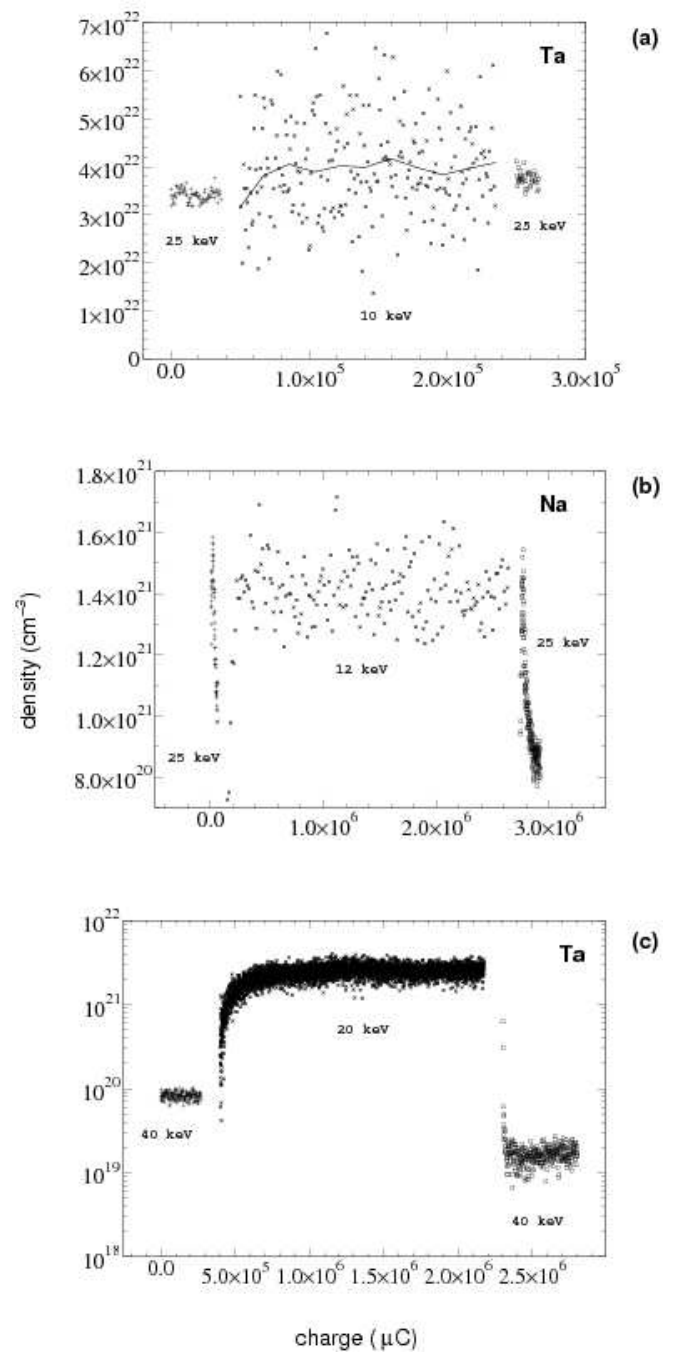}
\par\end{centering}

\caption{\label{fig:gegenbsps}Development of deuteron densities depicting
counter examples to the screening enhancement as in Fig.~\ref{fig:expprocedure}.
(a) A medium thick layer obliterates the screening enhancement discontinuities
at high densities, here at the example of an oxide layer on Ta. The
full line shows the progression from a refinement of the statistics
by recalculation with an increased stepsize. Targets featuring low
hydrogen binding ability hence allowing only for low and unstable
densities with quick profile shifts: (b) A thick metal oxide layer
overtopping the ion range on a metallic Na disk. (c) Heating vanquishs
the hydrogen metal bond, here beam heating of a $7\,\micro\meter$
Ta foil.}

\end{figure}

In order to limit the layer formation the totally implanted charge
was reduced \citep[Sec.\ 4.2]{nimb06}. For the analysis a more sophisticated
expression for the yield in (\ref{m:redy}, \ref{m:Fnorm}) was used
based on a model of the target with three stacked layers \citep[Sec.\ 4.3]{nimb06}:
The top layer consisting of either metal oxide or carbon, a deuterized
zone of the metal and the bulk of the metal containing essentially
no hydrogen. Each can have different thicknesses and relative deuterium
contents. The results for $U_{e}$ in Fig.\ref{fig:fituxim} were
obtained with only the additional parameter $\xi_{M}$ for the thickness
of the deuterated zone in the metal in energy equivalent units of
the stopping \citep[Sec.\ 5]{nimb06}. The differences for Ta-A and
Ta-E are already considerable though the thicknesses of the surface
layers were small and just started forming. Fig.~\ref{fig:emic}
shows the beginning of the formation of a carbon layer starting from
islands which will eventually cover the whole surface in concordance
with experiences from thin film technology \citep{ensinger97}. Ta-C
has already a relatively thick carbon layer which strongly reduced
the screening energy. Just as the metal oxide layer does in Ta-D.
Those layers were just thick enough in order to be included in the
model and infer their thickness. The thickness of the metal oxide
layer is $0.09\,\sqrt{\kilo\electronvolt}$, which conforms to about
$7\,\nano\meter$. The corresponding screening energy would be $433\,\electronvolt$
\citep[Sec.\ 5, Table\ 2]{nimb06}. $15\,\nano\meter$ are enough
to let the screening enhancement completely vanish \citep[Sec.\ 4.3]{nimb06}.
The deduced deuteron density is hardly affected and still in the vicinity
of the stoichiometric ratio as the example in Fig.~\ref{fig:gegenbsps}(a)
shows \citep[Sec.\ 6, Fig.\ 13.(e,f)]{nimb06}. %
 Much thinner surface layers already reduce the inferred screening
energy considerably. So the real value for the screening energy of
Ta is possibly around $400\,\electronvolt$ \citep[Sec.\ 5, Table\ 2]{nimb06}.
Anyhow, the screening energy values ranging from $210-460\,\electronvolt$
give an imagination of the systematic error originating from the surface
layer formation. Carbon can achieve high deuteron densities but it
does not show the electron screening effect as Fig.~\ref{fig:PdCF}
proves. Thin deuterated carbon layers can, however, simulate a screening
enhancement as inhomogeneous density profiles can do \citep[Sec.\ 4.3]{nimb06}.
Though the former could be excluded in our experiments but is a theoretical
possibility when thin deuterated carbon layers form on targets containing
few deuterium in segregation as below.

As already said, the metal oxide contains only few deuterium in segregation.
Those low densities are unstable and change under different conditions.
At the example of a Na target with a very thick metal oxide layer
the development of the calculated deuteron density is illustrated
in Fig.~\ref{fig:gegenbsps}(b) (also \citep[Fig.\ 13.(b-d)]{nimb06}).
The density estimates are calculated from the reduced yield as previously
described. Before the monitor measurement at $25\,\kilo\electronvolt$
a measurement at a low energy had been taken. Then the density quickly
decreased at $25\,\kilo\electronvolt$. Thereafter a measurement at
$12\,\kilo\electronvolt$ were started. Now, the density very quickly
increased reaching a higher level than at $25\,\kilo\electronvolt$.
But the discontinuity at the beginning was in the opposite direction.
The density for the sequencing monitor measurement started once again
at a high density which quickly decreased. The discontinuity at the
beginning was once again in the wrong direction. So there is definitely
no screening in contrast to the positive case of Fig.~\ref{fig:expprocedure}.
The quick shifts in the densities after the change of the implantation
energy going to a 'saturation' level originate from a shift of the
deuteron distribution depth profile in the metal oxide linked to the
different ranges of the ions \citep[Sec.\ 6, Fig.\ 14]{nimb06}. With
our method of recording a yield function $Y(q)$ over the implanted
charge we can recognize those shifts and reject them. If, however,
only the total yields of the long time measurements are regarded as
in the usually applied standard method (like in the other experiments
discussed in Sec.~\ref{sec:ddexps}) their comparison would erroneously
lead to a screening interpretation.

The same problem arises when working with low implantation densities
below the stoichiometric ratio even when the metal oxide layer is
negligible. Except for insufficient implantation the density remains
low if the thermal energy of the deuterons is higher than their chemical
binding energy to the metal so that they can float. This applies mainly
to transition metals with low ability to bind hydrogen (groups 6A-8A,
1B) or if the metals are heated. An example for the consequences of
heating is shown in Fig.~\ref{fig:gegenbsps}(c) for a Ta-foil of
$7\,\micro\meter$ which was heated by the beam power. One observes
the same behaviour and no real screening enhancement. The density
returns to an equal saturation level if the surrounding conditions
are the same, i.e.~same beam energy, current, target heat flow etc.
Tests with a Au-foil showed an alike behaviour. The most effective
heat transportation mechanism in solids is the free electron gas.
Cooling the target holder has little effect since the thermal resistance
at the connection is very high. Besides from heating the density profile
of the deuterons in target materials with low binding ability for
deuterons (metal oxides, metals with low affinity to hydrogen, metals
at high temperatures) is also changed by direct projectile hits and
close phonon generation at the target deuterons depending on the beam
energy. Furthermore, the metal oxide as a thermal insulator will be
considerably heated by the beam power. It is therefore preferable
to use thick target disks at moderate temperatures with high densities.
On the other side, cooling a target to very deep temperatures would
transform it into a cryogenic trap accumulating water in thick layers
on its surface prior to irradiation promoting the oxidation. The detailed
investigation is covered in \citep{dis}.

Summarizing, our data analysis method is independent of the absolute
deuteron density and allows for the discrimination between changes
in the reaction yield due to the density dynamics as in Fig.~\ref{fig:gegenbsps}(b,c)
which are rejected and actual changes in the cross section which become
manifest in the discontinuities at the edges of the measurements like
in Fig.~\ref{fig:expprocedure}. That the discontinuities signify
cross section modifications is further ensured by analyzing measurements
which are taken in proximity of the stoichiometric ratio only, where
changes of parameters like beam flux and temperature have marginal
influence on the overall deuteron density, at most. The error of $F_{\mathrm{norm}}$
is a convolution of the error from the counting statistics and long
term changes of the density. The use of high vacuum systems will inevitably
cause the build-up of contamination layers. Thanks to our analysis
method those layers can only diminish the inferred screening energy
since feigned enhancements due to density dynamics get rejected \citep[Sec.\ 6]{nimb06}.
The utilization of carbon hydrides embanks the layer formation enabling
the results in HV at all \citep[Sec.\ 4.2]{nimb06}. Indeed this is
a difficult and labile equilibrium in the residual gas. So layers
are present, which were examined by EDX allowing for a relative measurement
of element abundances \citep{dis}. But the thickness can hardly be
quantified because of the fractal structure of the target surfaces
(e.g.~Fig.~\ref{fig:emic}). Though the model suggests that $15\,\nano\meter$
are sufficient to completely dispose of the screening enhancement.
All in all, the obtained screening energies represent lower limits
to the real value. The magnitude of the dominating systematic error
from the unknown layer thickness can be assessed by the measurements
in Fig.~\ref{fig:fituxim}.

\section{\label{sec:results}Results}

\subsection{\label{sec:results.exp}Experiment}

The experimentally determined results for the screening energies are
summarized in Table \ref{tab:resultcampI}. %
\begin{table}
\caption{\label{tab:resultcampI}Screening energies}

\begin{ruledtabular}

\begin{centering}
\begin{tabular}{lll}
Metal & MD$_{x}$\footnotemark[1] & $U_{e}$ in \electronvolt\tabularnewline
\hline
Ta & 0.9 & 322 $\pm$ 15\tabularnewline
Zr & 2.1 & 297 $\pm$ 8\tabularnewline
Al & 0.8 & 190 $\pm$ 15\tabularnewline
Sr & 1.0 & 350 - 800\tabularnewline
Li & 0.03 & $\lesssim150$\tabularnewline
\hline 
Na & 0.03 & ---\footnotemark[3]\tabularnewline
Pd & 0.3\footnotemark[2] & 313 $\pm$ 2\tabularnewline
C & ---\footnotemark[4] & 0\tabularnewline
\end{tabular}
\par\end{centering}

\footnotetext[1]{Approximate average deuterium contents in relation to the number density of the metal}

\footnotetext[2]{The initial implantation was deliberately prematurely aborted.}

\footnotetext[3]{An oxidation layer impeded the determination of $U_{e}$}

\footnotetext[4]{Carbon density unknown. See text.}

\end{ruledtabular}
\end{table}
 The values from the campaign~I are in the upper part of the table.
In the lower part of the table are accessory results from the campaign~II.
In the second column of the table are the ratios of the deuterium
number density to that of the hostmetals. Since the deuteron density
can and does vary during a measurement these values are estimated
averages. The values for Strontium and Lithium are heavily impaired
by the layer formation due to the high reactivity of both metals,
for Li more than for Sr. Such expressed itself as strong variations
in the deuteron densites and accordingly in the reduced yields during
the course of the measurement, leading to ambiguous values for the
discontinuities of the reduced yields. So these screening energies
should be regarded as estimations, at best. The results were obtained
utilizing the equilibrium in the residual gas in order to keep the
target surface clean which was subsequently verified by EDX analysis
(Sec.~\ref{sec:expspecial} and more detailed \citep[Sec.\ 4.2]{nimb06}).

\begin{figure}
\begin{centering}
\includegraphics[width=1\columnwidth]{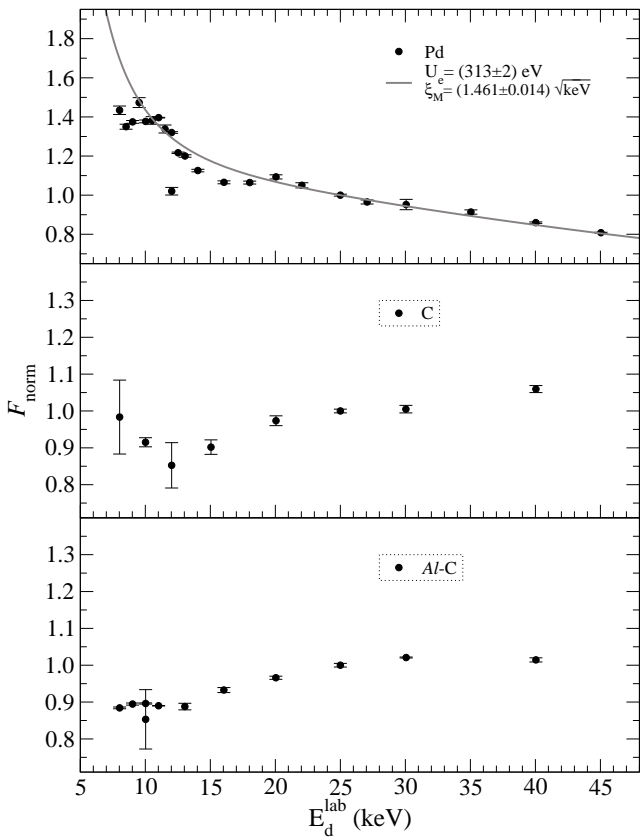}
\par\end{centering}

\caption{\label{fig:PdCF}Measured values of $F_{\mathrm{norm}}$ for Pd and
Carbon in two different compositions}

\end{figure}

The first plot in Fig.~\ref{fig:PdCF} is a measurement on palladium
with roughly equal residual gas conditions as for the Ta-measurements
\citep[Sec.\ 5]{nimb06}. The totally implanted charge was limited
for the same reason, i.e.~layer formation, as in the Ta-measurements
of Fig.~\ref{fig:fituxim}. %
 The beam spot contains traces of carbon specifically some dark stains
\citep[Fig.\ 8]{nimb06}. The other two plots are the experimental
prove that carbon has no screening enhancement. The drop to the lower
energies originated from a lower deuterium content in the upper layers
of the targets. This drop can also be caused by the voltage drop in
the plasma inside the RF-ion source \citep{kamke56} \citep[Sec.\ 2]{nimb06}
which has a higher impact for lower energies relative to the monitor
measurement at $25\,\kilo\electronvolt$. The two carbon targets were
preparated with different methods. The first one was made by deposition
of soot from a ethine flame on a backing plate. A flame of ethine
(C$_{2}$H$_{2}$) burning with insufficient oxygene supply produces
very pure carbon. However, the material is amorphous and rather fluffy.
Accordingly, the deuteron density reaches only values of about $1.5\cdot10^{22}\,\centi\meter\rpcubed$.
The second target is a carbon film produced by the irradiation of
aluminium with high decane pressure. That in such way deposited carbon
was compactified by the impacting beam ions while forcing it to adopt
the lattice structure of the substrate to a certain extend \citep{hering99}.
Hence the density of the carbon atoms is higher, so is the deuteron
density with about $5\cdot10^{22}\,\centi\meter\rpcubed$. Howbeit,
these are only estimates since the carbon densities are not known
and as a result the correct stopping power coefficient which is required
(\ref{m:redy}) \citep[Eq.\ (10)]{nimb06} neither. Anyway, the resulting
enhancement factors show no significant disagreement. Thus, carbon
films present no signs of electron screening. These results are listed
in the lower part of Table~\ref{tab:resultcampI}.

The highly reactive metal natrium corroded so easily that only low
deuteron densities could be achieved and no screening was visible.
Two tests with Y and Er led to thick metal oxide layers, too. Different
to the other metals the concomitant analysis of the $^{3}$He spectral
peak revealed in both experimental campaigns for Li, Na and Sr a significant
suppression of the neutron reaction channel and a simultaneous alteration
of the angular anisotropy \citep{NPAII06a,dis}.

\subsection{\label{sec:results.theo}Theory}

From the theoretical point of view the deuterized metals can be treated
as a strongly coupled plasma \citep{ichimaru93}. Since the velocity
of reacting nuclei is smaller than the Fermi velocity, the electron
screening effect corresponds to a static polarization of surrounding
conduction and bound electrons. Consequently, the electrostatic potential
energy between reacting nuclei of charges $Z_{1}$ and $Z_{2}$ shielded
in a metallic medium can be described within the self-consistent dielectric
function theory \citep{europhys04}: 

\begin{eqnarray}
V\left(r\right) & = & \frac{Z_{1}Z_{2}e^{2}}{r}\Phi\left(r\right)\nonumber \\
 & = & \frac{Z_{1}Z_{2}e^{2}}{\left(2\pi\right)^{3}}\int\frac{4\pi e\varphi_{1}\left(q\right)e\varphi_{2}\left(q\right)}{\varepsilon_{\nu}\left(q\right)\varepsilon_{c}\left(q\right)q^{2}}\exp\left(iqr\right)d^{3}q\nonumber \\
 &  & \stackrel{r\rightarrow0}{\longrightarrow}\frac{Z_{1}Z_{2}e^{2}}{r}-U_{pol}\label{eq:theo.pot}\end{eqnarray}
 The wave-number dependent dielectric functions $\varepsilon_{\nu}$
and $\varepsilon_{c}$ describe polarization of valence and core electrons
of host atoms induced by a charged impurity taking into account the
short range electron correlation and the exchange interaction between
electrons (for details see \citep{europhys04}). $\Phi\left(r\right)$
and $\varphi_{i}\left(q\right)$ functions are the screening function
and electronic charge-formfactors of reacting nuclei, respectively.
At small distances (applicable for nuclear reactions and decays) the
potential energy can be approximated using the energy independent
polarization screening energy $U_{p}$ which scales with the product
of the charges of the involved nuclei. For the d+d reactions we used
the self-consistent charge form-factor $\varphi\left(q\right)$ within
the Thomas-Fermi approximation \citep{grosso00,europhys04}:\begin{equation}
\varphi\left(q\right)=1-z+\frac{zq^{2}}{\left(q^{2}+k_{TF}^{2}\right)}\label{eq:theo.formfac}\end{equation}
 Here, the Thomas-Fermi wave number $k_{TF}^{2}=6\pi e^{2}n/E_{F}$
has been applied; $n$ and $E_{F}$ are the electron number density
and the Fermi energy, respectively. The number $z$ corresponds to
the fraction of electrons bound to deuterons and is for metals close
to unity. Since we are interested in the evaluation of the strongest
possible screening effect, we uniformly set $z=1$ for all target
materials. In the absence of screening $\varepsilon_{\nu}\equiv\varepsilon_{c}\equiv1$
and $z=0$, $V\left(r\right)$ reduces to the bare Coulomb potential
($\Phi\left(r\right)\equiv1$).

In the metallic lattice, besides electrons also positive ions can
contribute to the screening of the Coulomb barrier between reacting
nuclei. This effect, called cohesion screening \citep{europhys04},
can be calculated as a gain of the potential energy of two deuterons
in the lattice field of the host metal compared to that of the helium
atom produced in the fusion reaction. To calculate the potential energies
we used the universal ion-ion interaction given by Ziegler, Biersack
and Littmark \citep{ziegler85}. For a rough estimation of the cohesion
screening energy $U_{coh}$, we calculated the potential energy gain
resulting from the surrounding 12 host atoms assuming the same fcc
crystal structure for all target materials investigated. The cohesion
screening is a slowly increasing function of the atomic number. The
total screening energy is the sum of both contributions $U_{e}=U_{pol}+U_{coh}$.

\begin{figure}
\begin{centering}
\includegraphics[width=1\columnwidth]{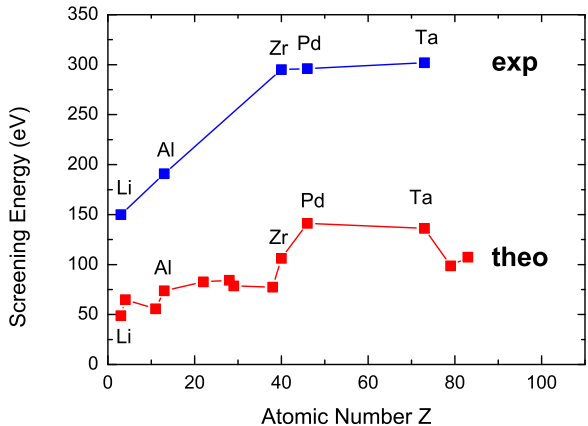}
\par\end{centering}

\caption{\label{fig:Zdep}(Color online) Comparison between experimental and
theoretical screening energies}

\end{figure}
\begin{figure}
\begin{centering}
\includegraphics[width=1\columnwidth]{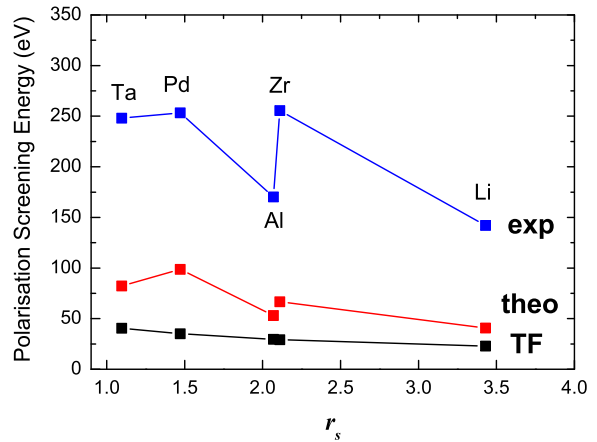}
\par\end{centering}

\caption{\label{fig:Rdep}(Color online) Experimental and theoretical polarization
screening energies versus the electron gas parameter $r_{S}$. For
comparison the Thomas-Fermi screening of the electron gas is presented.}

\end{figure}

 The results of the theoretical calculations obtained for the d+d
reactions taking place in different metallic targets are presented
in Fig.~\ref{fig:Zdep} together with our experimental values. The
electron screening energies moderately increase with the atomic number
of host atoms \citep{NPAII06b} reaching for heavier nuclei the value
of about $300\,\electronvolt$. The experimental target material dependence
agrees with the theoretical expectations. However, the experimental
screening energies are larger by a factor of about 2 compared to the
theoretical values. Since the experimental screening energies obtained
for insulating materials are much smaller ($<50\,\electronvolt$)
\citep{NPAII06b} and taking into account that the screening contributions
arising from polarization of bound host electrons and cohesion should
be similar for both metallic and insulating targets, we can conclude
that the enhanced screening effect results from conducting electrons.
Thus, for a comparison between different target materials the electron-gas
parameter $r_{S}=\left[3/\left(4\pi n\right)\right]^{\frac{1}{3}}/a_{0}$
where $n$ and $a_{0}$ are the valence electron density and the Bohr
radius, respectively, is much more suitable. Using this parameter,
the experimental polarization screening energies obtained by subtraction
of the theoretical cohesion contribution are displayed in Fig.~\ref{fig:Rdep}
together with the theoretical polarization screening energies. %
 Now, the quality of the theoretical description is much better visible.
In contrast to the simple Thomas-Fermi model \citep{grosso00} providing
for free electrons a smooth dependence of the screening energy given
by $U_{TF}=Z_{1}Z_{2}e^{2}\left[4/\left(\pi a_{0}\right)\right]^{\frac{1}{2}}\left(3\pi^{2}n\right)^{\frac{1}{6}}=2Z_{1}Z_{2}e^{2}\left[9/\left(4\pi^{2}\right)\right]^{\frac{1}{6}}r_{S}^{-\frac{1}{6}}$
the dielectric function theory describes fluctuations of the experimental
polarization screening energy very well. The fluctuations result from
the polarization of bound (core) electrons which contribution to the
total screening energy depends very strongly on their binding energy
\citep{europhys04}. If the bound electron contribution is eliminated
from the experimental polarization screening energies we get experimental
values for the free electron polarization which can be parameterized
by a smooth dependence on $r_{S}$ $U_{polf,f}=Z_{1}Z_{2}(250\pm20)\,\electronvolt/r_{S}^{\frac{1}{2}}$.
This result can be used for an estimation of the free electron contribution
in metallic environment to the screening energy in reactions between
nuclei with charges $Z_{1}$ and $Z_{2}$. Different to the d+d reactions,
the contribution coming from electrons bound by heavier reacting nuclei
is much larger and should be included separately. This can be calculated
as the gain in electron binding energies between distant atoms and
the final united atom. Similar results are to obtain using the Thomas-Fermi
model leading to $U_{e,b}\left(TF\right)=1.13Z_{1}Z_{2}e^{2}\left(Z_{1}^{\frac{1}{2}}+Z_{2}^{\frac{1}{2}}\right)^{\frac{2}{3}}/a_{0}$
\citep{genf06}. In the case of heavier nuclei the cohesion screening
energy can be neglected, since the strength of the interaction with
the lattice atoms increases much weaker than the product $Z_{1}Z_{2}$.
Thus, the total screening energy is only the sum of the free electron
and bound electron contributions. The same estimation can also be
applied for radioactive $\alpha$ and $\beta$ decays \citep{genf06}.

\begin{figure}
\includegraphics[width=1\columnwidth]{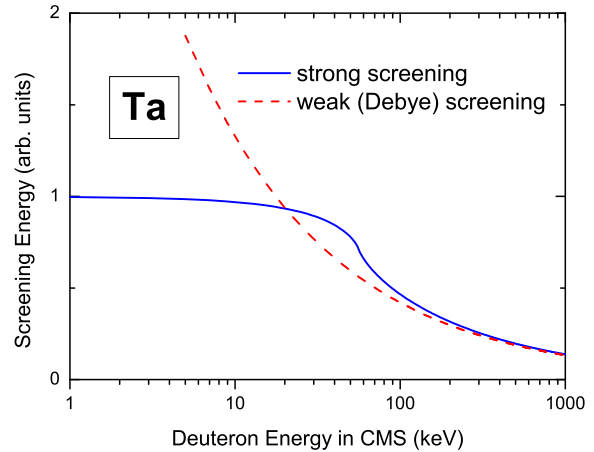}\caption{\label{fig:DynamTa}(Color online) Screening energy dependence on
the projectile energy. The Debye screening is applicable only for
deuteron energies larger than the Fermi energy ($56\,\kilo\electronvolt$
for Ta) or equivalently for plasma temperature larger than the Fermi
temperature ($1.8\cdot10^{5}\,\kelvin$ for Ta).}

\end{figure}

The dielectric function theory does not predict any temperature dependence
of the polarization screening energy unless the electron density of
the target material remains constant and the projectile velocity is
smaller than the Fermi velocity. That is typical for a strongly coupled
plasma. For velocities higher than the Fermi velocity the electrons
are not able to follow the ions and the electron screening gets weaker.
In this limit of a weakly coupled plasma (Debye-H\"{u}ckel limit)
the screening length becomes larger than the mean atomic distance
and classic description of the electron screening is applicable. The
screening energy is inversely proportional to the square root of the
kinetic energy or equivalently of the plasma temperature ($U_{e}\sim\frac{1}{\sqrt{E}}\sim\frac{1}{\sqrt{T}}$).
An analytical formula connecting both limits has been derived by Lifschitz
and Arista \citep{lifschitz98} for the stopping power of moving ions
in the electron gas and can be applied for the electron screening
in nuclear reactions \citep{NPAII06b}. Thus, the velocity dependence
of the screening energy can be given as follows: \begin{equation}
U_{dyn}^{2}=U_{ad}^{2}\left[\frac{1}{2}+\frac{v_{F}^{2}-v^{2}}{4v_{F}v}\ln\left|\frac{v+v_{F}}{v-v_{F}}\right|\right]\label{eq:veldep}\end{equation}
 where $U_{dyn}$ and $U_{ad}$ denote dynamic and adiabatic screening
energies, respectively. The Fermi velocity $v_{F}$ depends on the
electron density and therefore is characteristic for the target material.
The above relation calculated for the d+d reactions in the Ta environment
is presented in Fig.~\ref{fig:DynamTa}. %
 Additionally, the energy dependence of the Debye-H\"{u}ckel screening
is shown. It is visible that the electron screening can be described
by the Debye-H\"{u}ckel theory only for projectile energies higher
than the Fermi energy (the Fermi energy of deuterons in Ta amounts
to about $56\,\kilo\electronvolt$) or equivalently for temperatures
higher than the Fermi temperature (for Ta $\sim\:1.8\cdot10^{5}\,\kelvin$).
Thus, in the cases discussed here, the Debye-H\"{u}ckel screening
is not applicable for both nuclear reactions and radioactive decays.

\section{\label{sec:compare}Comparison with other experiments}

In view of the augmented information provided by our differential
analysis method and experimental procedure the results of other groups
will be discussed.

\subsection{\label{sec:ddexps}d+d Experiments}

\begin{figure*}
\begin{centering}
\includegraphics[width=0.9\textwidth]{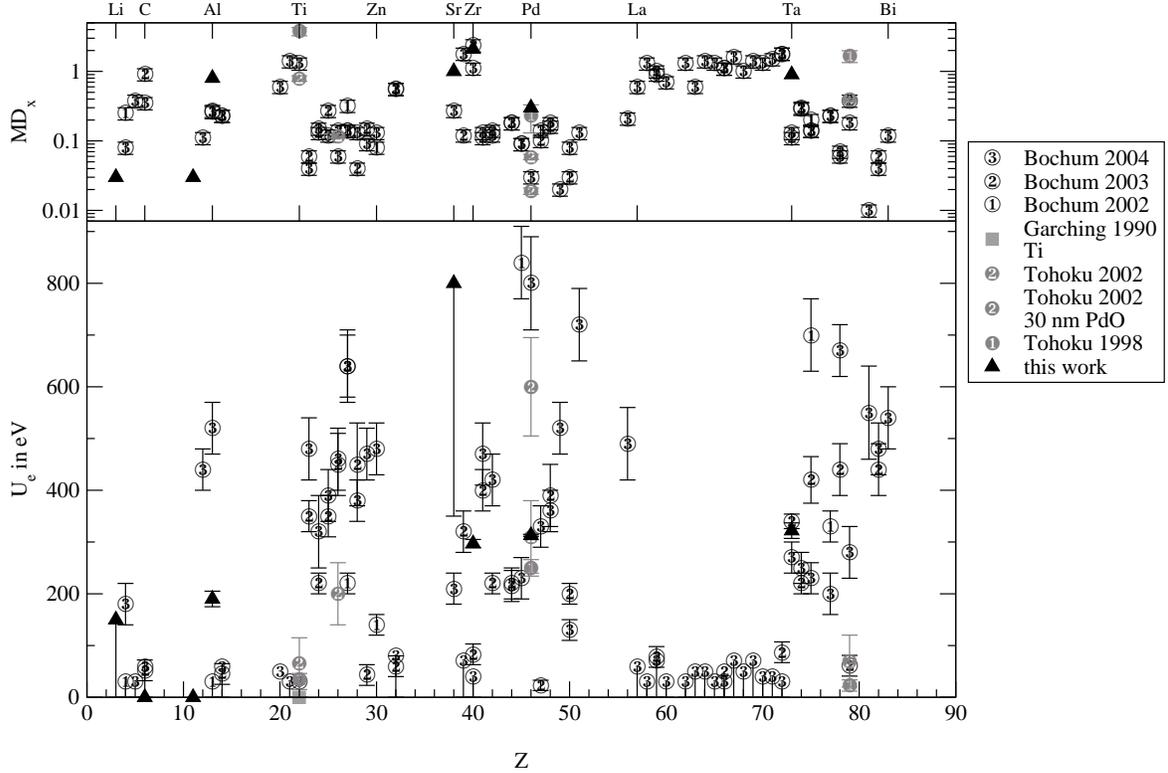}
\par\end{centering}

\caption{\label{fig:overv}Overview of screening experiment results. Legend:
Garching 1990 \citep{roth90}; Tohoku 1998 \citep{yuki98}, 2002 \citep{kasagi02};
Bochum 2002 \citep{rolfs02b}, 2003 \citep{rolfs03}, 2004 \citep{rolfs04}.
Bottom: Screening energies $U_{e}$. Top: Deuterium to metal ratio
$x$. The values for $x$ of \citep{kasagi02} were estimated from
Fig.~2 therein. The values of \citep{rolfs04} are the data base;
data points from \citep{rolfs03,rolfs02b} are included if they differ,
only.}

\end{figure*}

In Fig.~\ref{fig:overv} an overview of screening energy results
and appendant deuteron densities from other experiments is plotted.
 All were carried out in high vacuum systems hence suffering from
the same progressive oxidation process under ion irradiation with
the inherent problems as of Sec.~\ref{sec:expspecial} and \citep{nimb06}.
A quick glance already shows that the screening energy results are
pretty much scattered not revealing a pattern. But in conjunction
with the deuterium metal ratio aka the deuteron density peculiarities
become evident. Our high screening energy results (Table~\ref{tab:resultcampI})
were achieved at high absolute densities in the proximity of the chemical
stoichiometric ratio where the ion beam flux has no influence on the
target deuteron distribution. Whereas the high screening results of
the other groups were exclusively attained at low deuteron densities
$10^{-1\ldots-2}$ below the metal number density. Complementary high
densities did not yield enhanced screening in those experiments. Both
classes of screening results are associated to groups in the periodic
table exposing the chemical relationship with respect to the surface
reactions and hydrogen binding ability of the targets as described
in Sec.~\ref{sec:expspecial}. Particularly manifest is that in Fig.~\ref{fig:overv}
for the group 3A metals including the lanthanoides which have low
screening values at high densities and conform to the counter example
case of Fig.~\ref{fig:gegenbsps}(a). But also the transition metals
show three clusters of high screening results at low densities in
Fig.~\ref{fig:overv} corresponding to the case of Fig.~\ref{fig:gegenbsps}(b)
and Fig.~\ref{fig:gegenbsps}(c) respectively. Such will be substantiated
in the following.

\subsubsection{The Garching Experiment}

The first accelerator experiment with the aim to search for modifications
of the cross section in the d+d fusion reactions caused by the metallic
environment was done on Ti \citep{roth90}. No enhancement could be
observed. The measurements were performed on a $3\,\micro\meter$
Ti foil fixed in a copper target holder frame with flow channels for
LN$_{2}$ cooling and a thermocouple for temperature determination.
No effort was made to specify the deuteron density in the target.
Instead a fixed value from material research was adopted which is
inadequate. Due to the deep cooling of the target water is accumulated
on its surface, which produces under ion irradiation a considerable
oxide layer %
\footnote{This is sustained by \citep[Fig. 3]{roth90}. There the proton counting
number is plotted over the implanted charge with a coarse resolution.
The gradient of this curve is proportional to the yield and the deuteron
density (\ref{m:yieldexp}, \ref{m:yieldtheo}). The curve should
be linear at and after saturation. At $0.2\,\coulomb$ is a hump in
the curve after which the gradient is clearly lower than before. The
authors explain it by a hydrogen release from the foil because of
a short temperature increase. However the gradient remains significantly
below the previous value which altogether rather complies to the curve
from a metal oxide build up in \citep[Fig.\ 6]{nimb06}, C/O=0.4.%
}. In addition, the beam current up to $0.1\,\milli\ampere$ leads
to a distinct temperature increase inside the de-acceleration volume
of the ions in the thin foil which will also alter the density profile
away from the supposed unit value. All further measurements on Ti
in Fig.~\ref{fig:overv} resulted in very low screening values with
densities in proximity of the chemical stoichiometric ratio. The higher
the deuteron density, the lower the screening value. Ti is chemically
very similar to Zr, both belong to the group 4A. From our experience
Zr oxidizes readily. So a relatively thick metal oxide layer complying
to the third case in \citep[Fig.\ 13.(e,f)]{nimb06}, Fig.~\ref{fig:gegenbsps}(a)
explains the results.

\subsubsection{The Tohoku Experiments}

The results of \citep{yuki98,kasagi02} are based on the analysis
of the total yield of the proton measurements \citep[Eq. (1)]{yuki98}
with one detector at the lab-angle $90\degree$ and at the projectile
energy $E_{\mathrm{d}}\in\left[2.5,10\right]\kilo\electronvolt$ \begin{equation}
Y_{t}\left(E_{\mathrm{d}}\right)=\varepsilon N_{\mathrm{D}}\int\limits _{0}^{E_{\mathrm{d}}}\sigma\left(E\right)\left(\frac{dE}{dx}\right)^{-1}dE\label{eq:yuki98.1}\end{equation}
 after depth energy substitution (Sec. \ref{sec:setup}) with the
proton detection efficiency $\varepsilon$ a cross section parameterization
$\sigma$ of \citep{bosch92} and finally the target deuteron number
density $N_{\mathrm{D}}$ which is presumed to be constant for all
energies and ranges. With the stopping power relation the additional
error of the stopping power coefficients is introduced. In order to
determine and observe the density value repeated monitor measurements
were performed at $10\,\kilo\electronvolt$. The density was then
calculated from the yield $Y_{t}\left(10\,\kilo\electronvolt\right)$
by solving (\ref{eq:yuki98.1}) for $N_{\mathrm{D}}$ with the supposition
that the screening enhancement is there negligible. According to the
not unambigious text (\citep{kasagi02} and suitable back references)
for the quantification of the enhancement and extraction of the screening
energy the yields are normalized to the experimental one at $10\,\kilo\electronvolt$
thus becoming independent of the actual value of $N_{\mathrm{D}}$:
\begin{equation}
Y_{\mathrm{norm}}\left(\mathrm{E}\right)=\frac{f\left(E\right)\cdot Y_{t}\left(E\right)}{Y_{t}\left(10\,\kilo\electronvolt\right)}\label{eq:yuki98.ynorm}\end{equation}
 \begin{equation}
f\left(E\right)=\frac{Y\left(E\right)}{Y_{\mathrm{bare}}\left(E\right)}=\exp\left(\pi\eta\left(E_{\mathrm{CM}}\right)\frac{U_{e}}{E_{\mathrm{CM}}}\right)\label{eq:yuki98.f}\end{equation}
where the theoretical expression for the thick target enhancement
factor $f$ is simply adopted from the approximated term in (\ref{m:kontext.enhancedef-f})
of \citep{assenbaum87} for the enhancement of the cross sections
in thin targets. This expression is, however, derived for cross sections
based on an increase of the effective projectile energy. It might
be introduced in the integrant in (\ref{eq:yuki98.1}) at the most
and must not be pulled out of the integral in that manner. A thick
target enhancement factor should retain the energy integration like
in (\ref{m:Fscr}). The approximation as in (\ref{m:kontext.enhancedef-f})
is only valid for $E_{\mathrm{CM}}\gg U_{e}$ which is not longer
fulfilled by the given experimental circumstances with beam energies
of some \kilo\electronvolt and screening energies of several hundred
\electronvolt. This also means that the supposition of a negligible
screening enhancement at $10\,\kilo\electronvolt$ is not valid either.
Moreover, the term (\ref{eq:yuki98.f}) diverges for energies approaching
zero. While the invalid approximation leads to an underestimation
of the derived screening energy the neglect of the enhancement at
$10\,\kilo\electronvolt$ effectuates a gross exaggeration because
the curvature of the enhancement curve must be greater in order to
describe the steeper slope of the data (analog to the difference in
the curves \#1 and \#2 in \citep[Fig.\ 12]{nimb06}). The inclusion
of measurements taken at higher energies would have revealed such.
The deuteron density value obtained at $10\,\kilo\electronvolt$ ought
be heavily altered, too. The target holder was cooled with LN$_{2}$.
The constancy (and thence concluded the homogeneity) of the density
in the target was investigated by measurements at $10\,\kilo\electronvolt$
with target heating by different beam flux or a mounted heater in
the case of Pd in the interval $\left[170,230\right]\kelvin$ determined
with a thermocouple. The results in \citep[Fig.\ 1]{kasagi02} show
a strong dependency of the 'saturation' density on the target temperature
and material with a considerable general decrease with raising temperatures.
The density descends from Ti over Au, Fe, Pd to PdO. Conspicuous are
the differences in the deuteron densities between \citep{yuki98}
and \citep{kasagi02} for Ti, Au and Pd which are almost one order
of magnitude while the corresponding screening values accord within
their errors (Fig.~\ref{fig:overv}) though the latter $U_{e}$ are
generally higher. This discrepancy remains unexplained in \citep{kasagi02}.
While Au and Fe do not build up firm bonds to hydrogen the achieved
densities are proportionally higher at these deep temperatures. A
deliberately produced $30\,\nano\meter$ thick PdO layer on a Pd target
in \citep{kasagi02} yielded an especially high screening energy with
an especially low density. Such a thick PdO layer would show quick
shifts in the deuteron density profiles with higher averaged densities
at lower projectile energies like in the second case in \citep[Fig.\ 13.(b-d)]{nimb06},
Fig.~\ref{fig:gegenbsps}(b) when changing the projectile energy
and using the differential analysis method. So this large screening
is simulated by the density alteration during the total yield measurement.
Whereas the density in the Pd target of \citep{kasagi02} is noticeably
low which points on the formation of an oxide layer distinctly thicker
than on the Ti target approaching the PdO target. This means that
the screening value is also rather generated by density dynamics and
the agreement with our value is by chance. Albeit our Pd value was
obtained from measurements with a limited total ion dose and still
growing densities prior to saturation in order to minimize surface
contaminations the highest density of all experiments was achieved.
While the targets as described in \citep{kasagi02} are thick enough
$\sim1\,\milli\meter$ to guarantee an effective heat transport in
the bulk of the material by the electron gas, the heterogeneous target
Au/Pd/PdO with a total thickness of only $60\,\micro\meter$ (thereof
$0.1\micro\meter$ Au) \citep{yuki98} is too thin therefore leading
to a considerable temperature increase in the beam stopping volume
which is to this extent not detectable by any outside mounted thermocouple.
So, the observed high screening energy of $\left(602\pm23\right)\electronvolt$
can be explained by the shifts in the density profile due to elevated
temperatures like in Fig.~\ref{fig:gegenbsps}(c), \citep[Fig.\ 13.a]{nimb06}
and the heterogeneity of the target and accordingly the density.

In order to explain the ascertained relation between the screening
energy and the density depicted in \citep[Fig.\ 4]{kasagi02}, i.e.~high
screening comes along with low deuteron densities, the concept of
a deuteron 'fluidity' was introduced in \citep{kasagi02} where fluid
deuterons and conduction electrons are to behave like a hot plasma.
But in palladium oxide there are no conduction electrons. In view
of the stated density dynamics this explanation is decrepit. The explanation
by density dynamics is also sustained by the significantly larger
standard deviations of the repeated density measurements at $10\,\kilo\electronvolt$
for targets with low densities in \citep[Fig.\ 2]{kasagi02}. Indeed,
the saturation density in our experiments returns to the same level
for the same conditions but with higher deviations. In order to avoid
the observed temperature changes of the deuteron densities in the
targets the beam current was adapted in such a way that the power
input into the target was kept constant. Although in \citep{kasagi02}
is admitted that this procedure does not keep constant the power density
due to the stopping power relation the mobility of the deuterons is
not only influenced by the indirect ambient temperature but also by
direct ion interaction and changes in the distribution of the stopped
projectile deuterons. The authors conceded that they were not able
do detect possible short time changes in the proton counting rate.
With our differential method we did not observe any discontinuity
belonging to screening on oxidized targets with low absolute densities
being independent of the actual beam current and power input. The
trustiest screening value seems to be the result for Au in \citep{yuki98}
obtained at a high density which conforms to our test with a Au foil
at a very low density yielding no enhanced screening. But under ion
irradiation even noble metals can oxidize.

Postscript: Very recently the experiments have been continued using
the same accelerator set-up, procedure, and analyzing method as above
\citep{kasagi07}; consequently the same considerations apply. As
before the target holder was cooled with LN$_{2}$ further approaching
the boiling temperature of nitrogen. For the Sm target a screening
energy of $\left(520\pm56\right)\,\electronvolt$ was deduced %
\footnote{The experiment was meant to sustain the temperature dependence of
the inapplicable Debye model (Eq.~\ref{eq:rolfs.UeMod}), see next
subsection thereto.%
}. But not even an estimate for the deuteron density was given unlike
in the previous puplications. The high statistical errors in \citep[Fig.\ 5]{kasagi07}
and the vague statements regarding the deuteron 'fluidity' obtrude
the inference of a low density with the same consequences, too.

\subsubsection{The Bochum Experiments}

The largest data set of screening energies is provided by \citep{rolfs04,rolfs03,rolfs02b}.
The applied experimental procedure and data analysis method explained
in \citep{rolfs02,rolfs02b} is exactly the standard strategy in nuclear
astrophysics as described in \citep{rolfs88} including measuring
relative excitation functions with normalization to known cross sections.
As such it is a step back behind \citep{yuki98} where already concessions
to the special situation of hydrogen in metals were made. Again just
the total yield for the thick target of the measurement of the protons
with 4 detectors at the polar lab-angle of $\theta=130\degree$ at
the energy $E_{\mathrm{d}}\in\left[5,30\right]\kilo\electronvolt$
was determined. The total yield $Y\left(E_{\mathrm{d}},\theta\right)$
was repeatedly taken at fixed energies with equal stepsizes $\Delta$
of $0.5\,\kilo\electronvolt$ and $1.0\,\kilo\electronvolt$ for $E_{\mathrm{d}}>10\,\kilo\electronvolt$.
Thereof an energy differentiated yield $Y'(E_{\mathrm{d}},\theta)$
is calculated (\ref{eq:rolfs02.5}) in order to extract the cross
section (\ref{eq:rolfs02.7,8}) \citep[Eq.\ (5), (7), (8)]{rolfs02}%
\footnote{The symbol names granted by the authors have been changed for the
sake of the uniformity of the notation and comparability.%
}: \begin{eqnarray}
Y'(E_{\mathrm{d}},\theta) & = & (Y(E_{\mathrm{d}},\theta)-Y(E_{\mathrm{d}}-\Delta,\theta))/\Delta\label{eq:rolfs02.5}\\
 & = & \sigma(E_{\mathrm{eff}})\varepsilon_{\mathrm{eff}}(E_{\mathrm{d}})^{-1}\times\label{eq:rolfs02.7,8}\\
 &  & \underbrace{\Omega K_{\Omega}(E_{\mathrm{d}},\theta)W(E_{\mathrm{d}},\theta)}_{\alpha\stackrel{!}{=}\mathrm{const}.}\nonumber \end{eqnarray}
The energy integration prior to (\ref{eq:rolfs02.7,8}) vanished by
means of the mean value theorem of calculus leaving behind the integrand
to be evaluated at the effective energy $E_{\mathrm{eff}}\in\left(E_{\mathrm{d}}-\Delta,E_{\mathrm{d}}\right)$
where one half of the yield is attained. Except for $\sigma$ for
all other factors in (\ref{eq:rolfs02.7,8}) is assumed that their
change within the energy interval can be neglected. Moreover, the
angular terms are collected in the factor $\alpha$ which is supposed
to be constant for the whole energy range of the measurement. $\Omega$
is the solid angle of the detectors, $K_{\Omega}$ its transformation
to the CM-system and $W$ the angular distribution of the reaction
yield. $\varepsilon$ is there the stopping cross section, i.e.~the
energy loss per particle areal density $\electronvolt/\left(\mathrm{atoms}/\centi\meter\squared\right)$
and not the linear stopping power $\frac{dE}{dx}$ $\kilo\electronvolt/\micro\meter$.
The effective stopping cross section is assembled from the one for
deuterium and the host metal \citep[Eq.\ (9)]{rolfs02}: \begin{equation}
\varepsilon_{\mathrm{eff}}(E_{\mathrm{d}})=\varepsilon_{\mathrm{D}}(E_{\mathrm{d}})+x\varepsilon_{\mathrm{M}}(E_{\mathrm{d}})\label{eq:rolfs02.9}\end{equation}
 with the metal atom fraction M$_{x}$D. Thus, the dependence of the
composition of the target is completely shuffled from the stopping
factor into $x$. Consequently, the deuteron density described by
$x$ is forced to be fixed for all projectile energies, the range
in the target and the whole measurement series on the target. For
the determination of the absolute value of the cross section in (\ref{eq:rolfs02.7,8})
$x$ was scaled to a known cross section for gaseous deuterium \citep{greife95}
at $E_{\mathrm{d}}=30\,\kilo\electronvolt$. This means that the deduced
uniform deuteron density for the whole measurement series is only
dependent on the one value at $30\,\kilo\electronvolt$ and only there
validated, at most. So this method is even less sensitive to changes
of the density during the course of the experiment than \citep{yuki98}.
Then the S-factor is calculated. The screening energy is obtained
from another fit to the S-factor data with three parameters of the
expression in the second line of (\ref{m:kontext.enhancedef-f}) together
with a linear S-factor function. Furthermore, additional error sources
were introduced without need by sticking to the standard procedure:
repeated yield differences at fixed energies, introduction of the
effective energy, the stopping power coefficients, S-factor computation.
The errors of the computed S-factors are said to be dominated by the
spread in the yields $Y'$ from various runs (\citep[Sec.\ 4, p.\ 380]{rolfs02},
\citep[Sec.\ 2, p.\ 195]{rolfs02b}), i.e.~the yields were repeatedly
measured with stepwise increasing and decreasing beam energies. (This
implies that the errors from different $Y$-values are distinctly
higher than the corresponding statistical errors, which can be seen
in \citep[Fig.\ 1]{rolfs02b,rolfs04}.) As can be seen from the position
and errorbars of the datapoints in \citep[Fig.\ 1]{rolfs02b,rolfs04}
the differences of the $Y$-values must be significant. It complies
with our experience that the density profile returns to the same depth
averaged value for the same surrounding conditions but with higher
deviations at lower densities. The comparatively large errors relative
to the number of the datapoints from the non-linear fit routine for
the parameter $U_{e}$ reflect a significant correlation between the
3 fit parameters (as could have been read off the covariance matrix)
and hence judging the capableness of the applied model.

From Fig.~\ref{fig:overv} one can recognize once again the conspicuous
connection between the deuteron density and the screening energy like
in the data of \citep{yuki98,kasagi02}. High densities are linked
to low screening energies because of moderately thick metal oxide
layers as in the third case of \citep[Fig.\ 13.(e,f)]{nimb06}, Fig.~\ref{fig:gegenbsps}(a).
Examples are the elements of the groups 3A ($_{21}$Sc, $_{39}$Y
and the lanthanoides $Z=57-71$) and 4A (Ti, Zr and $_{72}$Hf) emphasizing
the chemical kinship with regard to the described surface reactions
in \citep[Sec.\ 4.1]{nimb06}, Sec.~\ref{sec:expspecial}. Low densities
generate high screening energy findings due to shifts in the density
profile either in thick metal oxide layers or materials with low hydrogen
binding ability as in \citep[Fig.\ 13.(a-d)]{nimb06}, Fig.~\ref{fig:gegenbsps}(b,c).
Such can be recognized at the transition metals (groups 6A-8A: $Z=24-28,\,42-46,\,74-78$)
for example. It is argued in \citep{rolfs02b} that the large enhancement
findings are most likely due to electron screening because the data
could be fitted well with the screening parameter $U_{e}$. In view
of the dispersion of the data points in \citep[Fig.\ 1]{rolfs02b,rolfs04}
their functional progression can also be described with the target
model of \citep[Sec.\ 4.3]{nimb06}. It implements a simple static
stepfunction for the density profile. The model can mimic a exponential
like increase towards low energies quite passable by an inhomogeneous
density profile with a super-deuterated surface layer alone without
screening enhancement, i.e.~$U_{e}\equiv0$, \citep[Fig.\ 10.c]{nimb06}.
A existing screening increase can also be largely exaggerated by the
density profile of a deuterated zone in the metal with a limited thickness
\citep[Fig.\ 10.d]{nimb06}. Those were only static density profiles.
A density profile dynamically changing with the energy as vindicated
by Fig.~\ref{fig:gegenbsps}, \citep[Fig.\ 13]{nimb06} could perfectly
imitate the exponential-like screening enhancement given the data
distribution. In contradistinction thereto our data does not allow
for such a description as quantitatively demonstrated in \citep[Sec.\ 5, Fig.\ 12]{nimb06}.
The Monte-Carlo code \noun{Srim }for the simulation of ion stopping
processes in matter was used in order to ratify the assumption of
a homogeneous depth distribution of the deuterons over the range of
the ions \citep{rolfs02}. But \noun{Srim} does not take into account
the ability of hydrogen to diffuse. The homogeneity assumption was
experimentally reconfirmed by a subsequent off-line ERDA on a $4\,\mega\volt$
tandem accelerator with the outcome that the distribution is uniform
within $10\%$ for 'most' materials \citep{rolfs02b}. Self-evidently
a subsequent examination cannot detect dynamic changes but only the
state of thermodynamic equilibration. Having been pointed to the problem
of oxidation \citep{EPS02b} RBS analysis was performed on the targets
with the result that there were 'no detectable surface contaminations'
with the exception of Al where there was an Al$_{2}$O$_{3}$ layer
with a thickness of about 150 monolayers \citep{rolfs03,rolfs04}.
Those findings prove that the resolution and sensitivity of the applied
analysis techniques are too low; at least the passivation oxide layers
from the unavoidable exposition to air with the used equipment should
have been visible. For both ERDA and RBS it is valid that light projectile
ions with a kinetic energy of some \mega\electronvolt\ cannot provide
a wide energy spectrum of the ejectiles which would be necessary in
order to resolve single atomic layers. Therefore a HIERDA with incident
energies of the heavy ions in the $0.1\,\giga\electronvolt$ order
of magnitude would be required with sophisticated magnetic analyzing
systems (e.g.~\citep{hering99}). This is additionally complicated
by the circumstance that these methods deliver expressive results
only if heterogeneous samples are made up of well defined layers.
This is not fulfilled for the implantation targets with indistinct
chemical composition and surfaces fractalized by embrittlement and
beam deterioration (Fig.~\ref{fig:emic}, \citep[Fig.\ 8]{nimb06}).
So the applied methods are not able to detect metal oxides with a
thickness of a few tens monolayers (some nanometers), which is already
sufficient to obliterate the screening enhancement (Sec.~\ref{sec:expspecial},
\citep[Sec.\ 4.3]{nimb06}) while they are not thick enough to affect
the applied density determination at $30\,\kilo\electronvolt$ significantly.

The thick oxide layer found on Al was defined to be of natural origin
due to the said property of Al to readily oxidize on air. Hence a
Kr ion sputtering treatment at $15$ or $35\,\kilo\electronvolt$
was applied prior to the implantation measurements in order to remove
those natural metal oxide layers which is the main difference from
\citep{rolfs02b} to \citep{rolfs03} and \citep{rolfs04}. This procedure
does not take into account that the major cause of the oxidation is
contributed by the water in HV systems under deuteron irradiation
which keeps going on nevertheless. While the high sputter yield of
the Kr ions may allow for a surface cleaning the large Kr atoms thoroughly
destroy the crystal structure of the target and get trapped in the
material fractalizing the surface and thus possibly even promoting
the oxidation process under subsequent deuteron irradiation since
the necessary annealing step is omitted. The deviations in the screening
energies between \citep{rolfs02b}, \citep{rolfs03} and \citep{rolfs04}
are in both directions, anyhow giving an indication for the magnitude
of the true error in the determination of the screening energies in
this way similar to our experiments on Ta (Fig.~\ref{fig:fituxim}).
Whether the increase or decrease of the screening finding comes from
an increase or decrease of the thickness of the oxide layer, or low
hydrogen binding ability of the metal, or a too thin overheated target
foil can scarcely be told afterwards on the basis of the available
information. But there are peculiarities. It becomes not clear which
beam currents were used, i.e.~54, 5 or 2.4 \micro\ampere, and how
they influence the stability and the inferred screening values. In
\citep{rolfs02b} it was reported about instable yields dependent
on the beam current for In and 'other elements with a low melting
point' %
\footnote{In \citep{rolfs04} In has a very high screening value without explanation
whether the problem has been solved or simply ignored, likewise the
elements Bi, Tl, Zn which have a low melting point. The elements Rh,
Re and Ir were measured with a beam current of $2.4\,\micro\ampere$
in \citep{rolfs02b} resulting in high screening energies which decreased
considerably in \citep{rolfs03,rolfs04}. Re decreased from $\left(700\pm70\right)\,\electronvolt$
over $\left(420\pm45\right)\,\electronvolt$ to $\left(230\pm30\right)\,\electronvolt$
indicating a beam current dependence even though these elements have
high melting points.%
}. The elements of the group 1B (Cu, Ag, Au) had a small screening
value in \citep{rolfs02b,rolfs03} complying to the gas target value
which became large in \citep{rolfs04}. That is in contradiction to
the very low screening energy for Au of \citep{yuki98} and our finding
of no screening. The extraordinary high screening value for Pd does
not change with the Kr sputtering but matches best to the PdO value
of \citep{kasagi02} which is another proof for the nevertheless continued
oxidation process. Due to the moderately thick metal oxide layer and
the stable deuteron density close to the stoichiometric ratio the
metals of the groups 3A, 4A including the lanthanoides neither allow
for a real screening observation nor a simulated screening by density
dynamics in the experiment of \citep{rolfs04}. In the coextensive
publications \citep{rolfs05b,rolfs06} these metals where heated to
$200\celsius$ thus overcoming the chemical bond between the metal
atoms and deuterium conveying it into segregation and leading to a
density drop of two orders of magnitude, i.e.~the case of \citep[Fig.\ 13.(e,f)]{nimb06},
Fig.~\ref{fig:gegenbsps}(a) is transformed to \citep[Fig.\ 13.(a-d)]{nimb06},
Fig.~\ref{fig:gegenbsps}(b,c). The than observed high screening
energies again can be informally explained by the density dynamics
due to the high mobility of the deuterons induced by the high temperature
and conjectural promoted metal oxide layer formation. This is made
clear at the example of Ti where five datapoints taken at different
temperatures show the transition in \citep[Fig. 3]{rolfs05b,rolfs06}.
In \citep{rolfs03} it was reported on difficulties in attaining stable
reaction yields for Ta at high temperatures which was subsequently
not further elucidated. The stability test for the density in \citep{rolfs05b}
is inapplicable since the used analysis method cannot recognize the
short time changes of the density.

The intention of these comprising experiments was to find a connection
between the observed screening energy and some electronic properties
of the elements, something that is to be underscored since it is an
important step towards the understanding of this phenomenon. The authors
propose the Hall coefficient to be this quantity stating that the
effective density $n_{\mathrm{eff}}$ of the free charge carriers,
i.e.~electrons and holes likewise, form a Debye sphere $R_{D}$ around
the deuterons and thus generate the screening potential \citep{rolfs03}:
\begin{eqnarray}
R_{D} & = & \sqrt{\frac{\varepsilon_{0}kT}{e^{2}n_{\mathrm{eff}}\rho_{a}}}\label{eq:rolfs03.Rd}\\
U_{e} & = & \frac{e^{2}}{4\pi\varepsilon_{0}}\frac{Z_{p}}{R_{D}}\label{eq:rolfs03.Ue}\\
 & = & \frac{e^{3}Z_{p}}{4\pi\varepsilon_{0}}\sqrt{\frac{\rho_{a}}{\varepsilon_{0}k}}\sqrt{\frac{n_{\mathrm{eff}}}{T}}\label{eq:rolfs.UeMod}\end{eqnarray}
 with $\rho_{a}$ the number density of the atoms $T$ the temperature
of the free electron gas and $Z_{p}$ the atomic number of the projectile.
The classical Debye screening is, however, not applicable for low
temperatures (electron energies below the Fermi energy) and dense
plasmas (solid states) where the quantum mechanical effects dominate
and the screening effect depends only on the charged particle density
and not on the temperature \citep{salpeter54,NPAII06b,genf06}. Additionally,
the motion of the bound electrons simulating the hole is not free
but governed by quantum mechanical tunneling between neighbor atoms.
The fact that the screening energy is vanishing for high deuteron
densities is explained by the assertion that these metal hydrides
are insulators. This is not right for the majority of the metal hydrides
which are metallicly or covalently bound and retain their metallic
properties. In fact, the electrons of the hydrogen are added to the
conduction band of the metal. The Baranowsky-curve of the electric
resistance of metal hydrides shows that the resistance at the chemical
stoichiometric ratio is even lower than for somewhat lower densities
and comparable to the metal \citep{mhydrides68}. Using a $^{3}$He
beam on a deuterated $_{78}$Pt target via the reaction d($^{3}$He,p)$^{4}$He
\citep{rolfs02b} a screening energy was inferred about twice as high
($\left(730\pm60\right)\,\electronvolt$ at $1-3\,\micro\ampere$)
as for the d beam ($\left(440\pm50\right)\,\electronvolt$) which
was regarded as a confirmation of the $Z_{p}$ dependency (\ref{eq:rolfs.UeMod})
of the Debye hypothesis \citep{rolfs03}. In \citep{rolfs04} however
the screening energies for $^{3}$He ($\left(680\pm60\right)\,\electronvolt$)
and d beams ($\left(670\pm50\right)\,\electronvolt$) at Pt became
equal without explanation. The inconsistency of the Pt-data also comprises
the measurements \citep{rolfs05b,rolfs06} for the verification of
the temperatur dependence (\ref{eq:rolfs.UeMod}) of $T^{-\frac{1}{2}}$.
With the exception of the room temperature data point the other four
data points are equal within their error interval. So the temperature
dependence is based on a single uncertain point. Furthermore, the
findings for the metals of the groups 3A and 4A are in contradiction
to it which cannot be resolved by the introduction of a highhanded
function \citep[Eq. (4)]{rolfs05b,rolfs06}.

In order to arrive at a more quantitative assessment of the hypotheses,
methods of statistical data description and analysis can be applied
(e.g.~\citep{stahel02}). Of the 58 examined elements in \citep[Table 1]{rolfs04}
the effective charge densities calculated from the Hall coefficient
$R_{H}$ are selectively specified for the 25 elements with high screening
values only, because the authors erroneously precondemned the others
to be insulators with zero charge carriers. Effective charge densities
for the elements In, Sn, Sb, Pb, Bi which do not fit in the explanation
scheme were also omitted %
\footnote{\label{fn:Hall}The Hall coefficients originate from \citep{hurd72}.
In \citep{rolfs03} the values for $n_{\mathrm{eff}}$ for Sn and
Pb in Table~1 were left out on the grounds that they were unreasonably
high (Table footnote 'f'). In \citep[Table\ 1]{rolfs04} the values
for In ($-82$), Sn ($-84$), Sb ($-0.09$), Pb ($21$), Bi ($-4\cdot10^{-4}$)
were omitted without vindication (the values for Sb and Bi are much
smaller than expected). Instead the Hall coefficient for Pd was remeasured
with a better fitting result giving reason to doubt other values for
the Hall coefficients. But no description of the measurement procedure
was given.%
}. It needs to be particularly pointed out that the authors impute
themselves a flat error of 20\% for the Hall coefficients but not
the original experimental uncertainties. Both have severe impact on
the interpretation. %
\begin{figure}
\begin{centering}
\includegraphics[width=1\columnwidth]{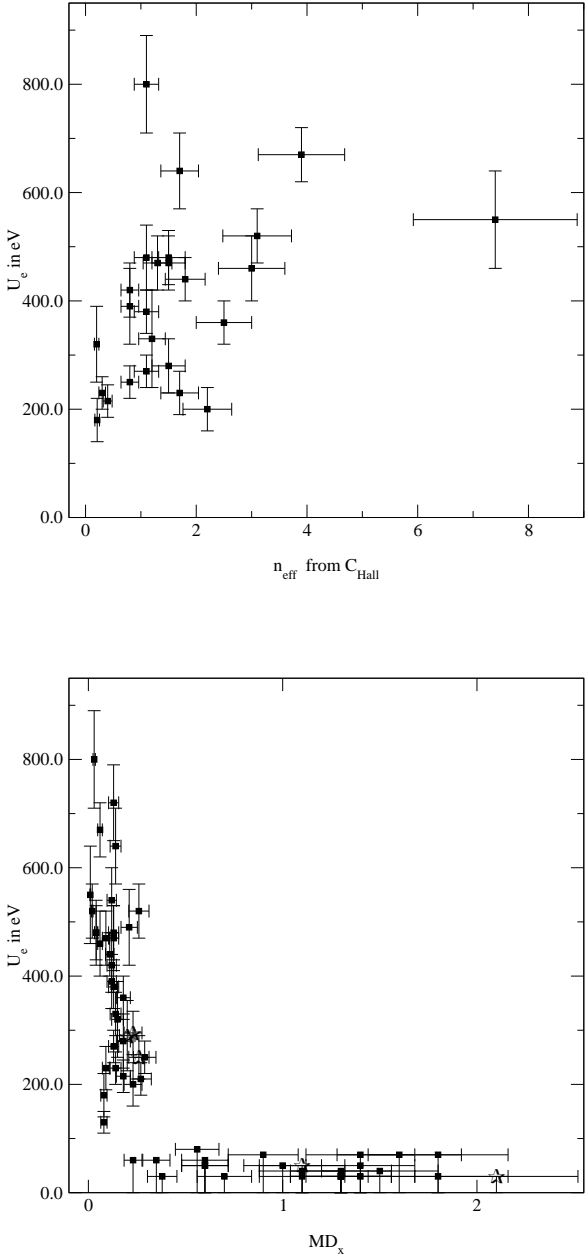}
\par\end{centering}

\caption{\label{fig:scatter}Scatter plots. Top: $U_{e}\leftrightarrow n_{\mathrm{eff}}$
with $n_{\mathrm{eff}}$ from the Hall coefficient $C_{\mathrm{Hall}}$.
Bottom: $U_{e}\leftrightarrow x$ with $x$ being the deuterium contents
in the compound MD$_{x}$. The star denotes temperature dependent
points for Ti.}

\end{figure}
 A visual survey based on Fig.~\ref{fig:scatter} already shows that
the distribution in the scatter plot for the case of $U_{e}\leftrightarrow n_{\mathrm{eff}}$
is rather dispersed while the distribution for $U_{e}\leftrightarrow x$
roughly indicates a hyperbolic connection. The temperature dependent
data points for Ti \citep[Table 1]{rolfs05b,rolfs06} are additionally
included in the bottom scatter plot (tagged with a star) demonstrating
the transition from a high stable density to a low instable density
allowing for the density dynamics which simulate the high screening
findings. Three testing methods for continuous variables were used:
Pearson's linear correlation $r$ which assumes a linear association
between the variables. The Spearman rank correlation $r_{s}$ measures
the monotone association between the variables and is therefore invariant
under monotone transformations. Kendall's $\tau$ is even more nonparametric
since it uses only the relative ordering of the ranks by counting
the inversions in the paired data points. It also enables the easy
inclusion of errors by adaptive binning. The latter both are robust
in opposition to the linear correlation. All such tests attempt to
falsify the null hypothesis of no correlation. Their correlation coefficient
describes the strength of the correlation ranging in $\left[-1,1\right]$
where 0 stands for no correlation and (-)1 for total (anti)correlation.
Complementary the $P$-value determines the significance of the obtained
correlation, the lower $P$ the higher is the significance. The results
are in tendency congruent wherefore representatives from the Spearman
rank correlation are listed in Table~\ref{tab:correls}. %
\begin{table}
\caption{\label{tab:correls}Spearman rank correlation tests}

\begin{ruledtabular}

\begin{centering}
\begin{tabular}{lld@{}l}
No.\footnotemark[1] & correl. & \multicolumn{1}{c}{$r_{s}$} & $P$-value\tabularnewline
\hline
1 & $U_{e}\leftrightarrow n_{\mathrm{eff}}$ & 0.4894 & 0.0130\tabularnewline
2 & $U_{e}\leftrightarrow x$ & -0.7997 & 5.1404$\cdot10^{-14}$\tabularnewline
3\footnotemark[2] & $U_{e}\leftrightarrow n_{\mathrm{eff}}$ & 0.2830 & 0.0768\tabularnewline
4\footnotemark[3] &  & 0.1240 & 0.4171\tabularnewline
5\footnotemark[4] &  & 0.0174 & 0.9096\tabularnewline
6\footnotemark[2] & $U_{e}\leftrightarrow x$ & -0.7466 & 2.3602$\cdot10^{-16}$\tabularnewline
\end{tabular}
\par\end{centering}

\footnotetext[1]{Data of \cite[Table 1]{rolfs04}.}
\footnotetext[2]{Including \cite[Table 1]{rolfs05b,rolfs06}.}
\footnotetext[3]{Including \cite[Table 1]{rolfs05b,rolfs06} and $n_{\mathrm{eff}}$ for the omitted elements.}
\footnotetext[4]{Additional consideration of the sign of the Hall coefficient.}

\end{ruledtabular}
\end{table}
 The assumed functional dependency in (\ref{eq:rolfs.UeMod}) is tested
on the restricted data set \citep[Table 1]{rolfs04} (\#1); the correlation
coefficient is below $0.5$ with a low significance. In contradistinction
thereto the correlation to the density $x$ has considerably higher
values with utmost strong significances (\#2). Including the temperature
dependent data \citep[Table 1]{rolfs05b,rolfs06} into the calculations
leads to a considerable decrease in the correlation coefficient (\#3)
which is further reduced when $n_{\mathrm{eff}}$ for the omitted
elements In, Sn, Sb, Pb, Bi is regarded (\#4), with a concurrent decrease
in the significance. The consideration of the sign of the Hall coefficient
lets the correlation approach zero (\#5). On the other hand the enlarged
data set has only slight impact on the correlation to $x$ (\#6).
Logarithmizing (\ref{eq:rolfs.UeMod}) leads to a linear model with
a slope of $b=\frac{1}{2}$ and a positive intercept $a$ containing
the other quantities. Regression attempts based on it are listed in
Table~\ref{tab:regress}, $\sigma_{i}$ is here the corresponding
standard error of the two parameters. %
\begin{table}
\caption{\label{tab:regress}Regressions on $\ln n_{\mathrm{eff}}\mapsto\ln U_{e}$}

\begin{ruledtabular}

\begin{centering}
\begin{tabular}{lccccdc}
No.\footnotemark[1] & $a$ & $b$ & $\sigma_{a}$ & $\sigma_{b}$ & \multicolumn{1}{c}{$\chi^{2}$} & $Q$-value\tabularnewline
\hline 
1 & 5.87 & 0.41 & 0.03 & 0.04 & 110.6 & 1.99$\cdot10^{-13}$\tabularnewline
2\footnotemark[2] & 5.91 & 0.34 & 0.03 & 0.04 & 136.0 & 5.01$\cdot10^{-18}$\tabularnewline
3\footnotemark[3] & 6.03 & -0.02 & 0.02 & 0.01 & 385.3 & 1.50$\cdot10^{-56}$\tabularnewline
\end{tabular}
\par\end{centering}

\footnotetext[1]{Data of \cite[Table 1]{rolfs04}.}
\footnotetext[2]{$\delta n_{\mathrm{eff}}=10\%$}
\footnotetext[3]{Including \cite[Table 1]{rolfs05b,rolfs06} and $n_{\mathrm{eff}}$ for the omitted elements.}

\end{ruledtabular}
\end{table}
 The influence of the error $\delta n_{\mathrm{eff}}$ can be seen
in comparison of \#1 and \#2, where \#1 has the willfully by the authors
imputed error of 20\% while \#2 adopts a more realistic value of 10\%
in contrast. Anyway, the resulting slopes definitely stay behind the
necessary value of $\frac{1}{2}$. Making worse the $Q$-value which
is the goodness of the fit remains tiny. Conventionally if the goodness
is smaller than $10^{-3}$ the model is considered incorrect or the
errors are still roughly underestimated. Here (\#3) the inclusion
of the temperature dependent data \citep[Table 1]{rolfs05b,rolfs06}
and $n_{\mathrm{eff}}$ for the omitted elements In, Sn, Sb, Pb, Bi
into the calculations leads to a slope close to zero and a goodness
disqualifying the linear model. The value of the Hall coefficient
of Pd was doubted in \citep{rolfs04} and replaced by an own measurement
with minor impact. So in both instances, correlation and regression,
the explanation by the Debye hypothesis is ruled out. The disaffirmation
of the Debye hypothesis on basis of the correlation tests alone might
be disputable, together with the other aforementioned points it is
decrepeted. The density hypothesis as an alternative clearly could
not be falsified. Complemented by the preceding argumentation and
the physico-chemical effects as described in \citep[Sec.\ 4, 6]{nimb06}and
Sec.~\ref{sec:expspecial} the deuteron density dynamics provide
the explanation for the alleged screening results. Thus the assumptions
for reactions with heavier nuclei and radioactive decay are refuted
as well with the consequence that any experimental evidence offered
for them need to come under scrutiny, Sec.~\ref{sec:heavynuc}, \ref{sec:decay}.

\subsection{\label{sec:heavynuc}Experiments with heavier nuclei}

As a consequence of Sec.~\ref{sec:ddexps} the erratic high screening
findings of the other groups cannot really serve as a confirmation
of our results. Therefore the experiments on Li target nuclei achieve
special significance as a independent reassurance for the overall
effect of the enhancement of nuclear reactions in metallic environments.
With regard to the reliability the most outstanding result discussed
in Sec.~\ref{sec:compare} is from the experiment d+Li in binary
alloys with Pd and Au \citep{kasagi04}. The $\alpha$-yield from
the reactions at $^{6,7}$Li (natural abundance: 7.42\% $^{6}$Li,
92.58\% $^{7}$Li) in both alloys was observed normalized to the yield
at $75\,\kilo\electronvolt$ in complete analogy to the procedure
for the d+d reaction \citep{yuki98,kasagi02} and set in relation
to LiF targets. The inferred screening energies are $(1500\pm310)\,\electronvolt$
for PdLi$_{x}$ and $(60\pm150)\,\electronvolt$ for AuLi$_{x}$,
where $x$ was initially at 5-10\% for the alloys. The screening energy
for PdLi$_{x}$ is therewith one order of magnitude higher than our
value of $(190\pm50)\,\electronvolt$ for $^{6}$LiF which is in agreement
with the simple theory \citep{prev97}. The latter screening energy
is substantially smaller than the $(380\pm250)\,\electronvolt$ from
\citep{engstler92}. This is because a significant share of the increase
of the observed S-factor towards low energies is caused by a $2^{+}$subthreshold
resonance at $E_{x}\left(^{8}\mathrm{Be}\right)=22.2\,\mega\electronvolt$.
It needs to be included in the data analysis by sophisticated nuclear
reaction theoretical calculations \citep{prev97,rupr04}. The advantage
of this experiment on Li targets over deuterium is that the small
hydrogen atoms have a mobility in metals which is several orders of
magnitude greater than other atom species. Thus, different to the
auto implanted deuterium targets there is no such fatal target atom
density dynamics possible as in \citep[Fig.\ 13.(a-d)]{nimb06}, Fig.~\ref{fig:gegenbsps}(b,c).
The also here inevitable oxidation process will only decrease -- but
not completely supplant -- the Li fraction in the surface layers.
This explains together with the higher sputtering yield for light
atoms the observed asymptotic bisection of the yield with the ion
dose \citep[Fig. 2]{kasagi04} at the monitor energy of $75\,\kilo\electronvolt$.
Both leads to an inhomogeneous depth distribution of the Li target
atoms with a lower Li fraction at the surface and due to the higher
beam energies $\geq30\,\kilo\electronvolt$ with lower impact. So
this time the observed enhancement can be regarded as a lower limit,
too. Whereas the inferred value of the screening energy needs to be
corrected for the influence of the subthreshold resonance and the
same questions regarding the screening energy calculation of the authors
apply as in the case of deuterium. The low value for AuLi$_{x}$ is
so far in conformity with the negative findings for the d+d reaction
in Au \citep{rolfs02b,rolfs03,yuki98,kasagi02}. A similar experiment
\citep{rolfs05c} was later performed using the proton induced reactions
on $^{6,7}$Li in an environment of Li$_{2}$WO$_{4}$, Li-metal and
PdLi$_{x}$ ($x=1\%,0.01\%$). The results for the screening energies
of the reaction $^{7}$Li(p,$\alpha$)$\alpha$ are $(185\pm150)\,\electronvolt$
for Li$_{2}$WO$_{4}$, $(1280\pm60)\,\electronvolt$ for the metal
and $(3790\pm330)\,\electronvolt$ for PdLi$_{1\%}$ which were obtained
using standard procedures (\ref{eq:rolfs02.5}), (\ref{eq:rolfs02.7,8}),
\citep{rolfs02,rolfs88}. The results for LiPd$_{0.01\%}$ and the
reaction $^{6}$Li(p,$\alpha$)$^{3}$He agree within $1\sigma$.
In a microscopic view it is universally valid that the screening effect
depends on the impact of the electronic configuration of the environment
on the Coulomb barrier of the entrance channel only (e.g.~\citep{shoppa93,shoppa96}),
i.e.~the pure Coulomb energy is modified by a Yukawa factor for simplicity
$W(r)=\frac{1}{4\pi\varepsilon_{0}}\frac{Z_{p}Z_{t}e^{2}}{r}e^{-\frac{r}{\lambda_{\mathrm{A}}}}$with
$\lambda_{\mathrm{A}}$ being the screening length. As such the inferred
screening energy is merely the second term in a Taylor-expansion of
$W(r)$, i.e.~$U_{e}=\frac{1}{4\pi\varepsilon_{0}}\frac{Z_{p}Z_{t}e^{2}}{\lambda_{\mathrm{A}}}$,
and a coarse mathematical parametrization in the simple model \citep{assenbaum87},
(\ref{m:kontext.enhancedef-f}), \citep[Eq.\ (20)]{nimb06}. The screening
modification of the Coulomb potential only acts as if the projectile
gained $U_{e}$. So there is no 'acceleration mechanism' in reality
and one must neither decompose the screening effect nor transfer the
result of one environment to another as in \citep{rolfs05c} where
the 'atomic' screening energy for $^{7}$Li$\rightarrow$H$_{2}$
is used as a linear addend in the screening energy for p$\rightarrow$(Li-metal
or PdLi$_{x}$). Consequently, the screening energy is independent
of the isotopes in the reaction and should be equal for $^{1,2}$H+$^{6,7}$Li
in \citep{kasagi04} and \citep{rolfs05c}. Whereas there are two
discrepancies between \citep{rolfs05c} and \citep{kasagi04}: First,
the more than twice as high screening energy for PdLi$_{x}$ of \citep{rolfs05c}
relative to \citep{kasagi04}. But 4 of the 7 datapoints lay offside
the fitted curve and only the fit error is given for $U_{e}$ \citep[Fig.\ 1,2]{rolfs05c}.
Second the assertion in \citep{rolfs05c} that the yield remained
stable better than 10\% while \citep{kasagi04} observed a bisection
of the yield at $6\,\coulomb$ which is plausible due to irradiation
effects. Both discrepancies can be explained by the different target
fabrication techniques. In \citep{kasagi04} Pd and Li are made into
an alloy by arc melting while in \citep{rolfs05c} Li was inserted
in a Pd-disk in a plasma discharge. The latter is prone to depth inhomogeneities.
This was verified by a NRA analysis of the target using the $E_{\alpha}=958\,\kilo\electronvolt$
resonance with a width of $\Gamma=4\,\kilo\electronvolt$ in the reaction
$^{7}$Li($\alpha$,$\gamma$)$^{11}$B yielding the ascertainment
of a homogeneous depth distribution \citep{rolfs05c}. However, the
depth resolution of this method is limited by the energy uncertainty
and spread of the beam and the width of the resonance. The most prominent
example of the NRA is the $E_{\mathrm{N}}=6.385\,\mega\electronvolt$
resonance with a width of $\Gamma=1.8\,\kilo\electronvolt$ in the
$^{1}$H($^{15}$N,$\alpha$$\gamma$)$^{12}$C for the investigation
of hydrogen distributions. It has a minimal resolution ranging $5-15\,\nano\meter$
\citep{schatz92}. So the resolution of the Li-NRA is worse given
a 2.2 times higher width of the resonance. Since most of the yield
is contributed by the topmost atomic layers here too (\citep[Sec.\ 4.3, Fig.\ 10.d]{nimb06}),
an enlarged Li contents below the NRA-resolution at the surface would
explain the more than two times higher screening energy and the much
lower decrease of the yield with the ion dose. The high screening
value for PdLi$_{x}$ was regarded as a confirmation for the Debye
model. If this was true the measurements for AuLi$_{x}$ \citep{kasagi04}
should also have yielded a high value and not one close to zero, since
the d+d screening energy for Au of \citep{rolfs04} was about $280\,\electronvolt$
in difference to \citep{rolfs02b,rolfs03,yuki98,kasagi02} and our
observation.

The theoretical model of the electron screening presented in Sec.~\ref{sec:results.theo}
predicts different screening energies for different target material
environments. In the case of an insulator the electron screening should
reach the value of $190\,\electronvolt$, which results only from
the gain of the electron binding energies. For metallic environments
the contribution coming from free electrons has to be included additionally.
Due to different electron densities for Pd ($r_{S}=1.4$) and Li ($r_{S}=3.4$)
the free electron contributions to the screening energy is equal to
$660\,\electronvolt$ and $420\,\electronvolt$, respectively. Thus,
we finally expect total screening energies of $190\,\electronvolt$
for an insulating target material, $610\,\electronvolt$ for metallic
Lithium targets and $850\,\electronvolt$ for the PdLi$_{x}$ alloy.
Experimental results, despite large uncertainties, confirm different
electron screening energies for insulating and metallic materials
with various electron densities.

Extending this thread, a first effort was undertaken in \citep{rolfs06b}
to study the environmental influence for heavy nuclei using the (p,n)
reaction on $^{50}$V and $^{176}$Lu nuclei in an oxide, as pure
metal, and as an alloy with Pd in the energy range $0.75-1.5\,\mega\electronvolt$.
Because of insufficient cross section data the screening energies
were obtained by comparison with the metal oxides VO$_{2}$ and Lu$_{2}$O$_{3}$.
The inferred screening energies are $(27\pm9)\,\kilo\electronvolt$
and $(33\pm11)\,\kilo\electronvolt$ for V and PdV$_{10\%}$ and $(32\pm2)\,\kilo\electronvolt$
and $(33\pm2)\,\kilo\electronvolt$ for Lu and PdLu$_{10\%}$. The
comparison was done by taking the ratio of the yields between the
metal and the oxide alias the insulator \begin{equation}
R(E_{\mathrm{p}})=\frac{Y_{\mathrm{m}}(E)}{Y_{\mathrm{i}}(E)}=\frac{\int_{0}^{E_{\mathrm{p}}}\delta_{\mathrm{n}}(E)\varepsilon_{\mathrm{m}}^{-1}(E)\sigma_{\mathrm{m}}(E)dE}{\int_{0}^{E_{\mathrm{p}}}\delta_{\mathrm{n}}(E)\varepsilon_{\mathrm{i}}^{-1}(E)\sigma_{\mathrm{i}}(E)dE}\label{eq:rolfs06b.RY}\end{equation}
where $\delta_{\mathrm{n}}$ is the efficiency of the neutron detector.
It is now assumed that the ratio of the stopping cross sections between
the two materials can be expressed by an energy independent constant
$\alpha=\varepsilon_{\mathrm{i}}(E)/\varepsilon_{\mathrm{m}}(E)$
which is mathematically doubtful considering Bragg's rule \citep{bragg05}.
So the following substitutions where done $\varepsilon_{\mathrm{m}}(E)=\alpha^{-1}\varepsilon_{\mathrm{i}}(E)$
and $\sigma_{\mathrm{m}}(E)=f(E)\sigma_{\mathrm{i}}(E)$ with the
enhancement factor $f$ as in (\ref{m:kontext.enhancedef-f}) with
the presupposition of a constant $S$. The ratio of the integrals
is further simplified by the energy differentiation of the yields
using the effective energy as in (\ref{eq:rolfs02.5}), (\ref{eq:rolfs02.7,8})
\citep[Eq.\ (5), (7), (8)]{rolfs02} arriving at $R(E_{\mathrm{p}})=\alpha f(E_{\mathrm{eff}})$.
The screening energy resulted together with $\alpha$ from a fit to
the yield ratios. This procedure was, however, only applied to V.
The screening energies for Lu were gathered from the shift of the
Lewis peak along the energy axis between the different targets originating
from a narrow resonance close to $E_{p}=0.8\,\mega\electronvolt$
\citep[Fig.\ 3]{rolfs06b}. The Lewis effect comes from the discrete
energy loss of the projectiles in the target \citep{lewis62}. This
energy shift was indeed erroneously interpreted as the screening energy.
As already pointed out the screening effect is merely a modification
of the Coulomb barrier and no real energy shift. So this shift can
not originate from the screening effect. Thus, it is probable that
the energy shift is caused by target properties. Strikingly the oxide
targets being the normalization standard are made by pressing a metal
oxide powder into a cylindrical hole of a Cu disk. It is well known
from powder metallurgy and silicate technology that pressing of a
powder like in this case is insufficient in order to remove the hollow
spaces between the powder particles unless a sintering step is performed.
So the used metal oxide targets contain hollow spaces with a size
of the same order of magnitude as the powder particles. Consequently
the stopping of \mega\electronvolt\ protons is heavily altered in
comparison to a monolithic metal oxide ceramic and different in its
mathematical description to \citep{ziegler77}. We observed effects
of porous targets on the stopping \citep[Fig.\ 2]{NPAII06a}. So the
shift of the lewis peak can be explained by the differences of the
stopping between the porous metal oxide target and the metal targets.
Additionally, it is well known that the position and form of the Lewis
peak depends very critically on the composition, homogeneity and contamination
of the target \citep{lewis62,walter61,donhowe67} also \citep{rolfs88}.
This in turn casts serious doubts on the results for V. A critical
point of the data analysis (\ref{eq:rolfs06b.RY}) is the presupposition
that the ratio of the stopping cross sections of the metal oxide and
the metal is a constant over the energy. This is inappropriate for
the porous target and can lead to a misinterpretation of the data.
The conspicuously high errors of the screening and $\alpha$ values
from the fit --- about 33\% making the effect compatible with zero
within $3\sigma$ --- are a strong indication for a high correlation
between the two fit parameters showing the improperness of the fit
model. The covariance matrix of the fit parameters could have given
information about this.

From the theoretical point of view the large screening energies obtained
for the d+d reactions at energies below $20\,\kilo\electronvolt$
cannot be used for the estimation of the screening energies in the
above case since the proton energy is much higher and does not fulfill
the adiabatic approximation. Since the surrounding electrons are much
slower than the protons, the resulting screening energy obeys rather
assumptions of the sudden approximation and thus should be of order
of a few \kilo\electronvolt\ in contradiction to statements involved
in \citep{rolfs05c,rolfs06b}.

\subsection{\label{sec:decay}Radioactive decay of embedded nuclei}

As the electron screening enhances the cross section at low impact
energies, a similar effect can be expected for the radioactive decay.
However, since the energies of the decay products are fixed by the
Q value, only a few nuclei with lowest-energy emitters are candidates
for a measurable change in the lifetime. In general, for positive
charged ejectiles ($\alpha$ and $\beta^{+}$ decay), screening reduces
the Coulomb barrier and therefore enhances the decay rate while the
opposite is true for $\beta^{-}$ decay. As recently pointed out by
Zinner \citep{zinner07} the effect of a changed Coulomb barrier is
partially canceled by a modified Q value that stems from the extension
of the screened potential into the inner part of the nucleus. For
heavy nuclei the effect can still be strong as the screening potential
scales approximately with the product of the charge number of the
end nuclei.

Recently, based on an extrapolation of the Debye-H\"{u}ckel electron
screening model to low temperatures, it has been suggested that half-lives
of radioactive isotopes may change by orders of magnitude if they
are embedded in a metal lattice and cooled to cryogenic temperatures
\citep{rolfs_public2,rolfs_public3,rolfs06c,rolfs06b,rolfs05b} %
\footnote{This effect has also been proposed as a new method of disposing radioactive
waste from nuclear power plants \citep{rolfs_public2,rolfs_public3,rolfs06c}.%
}. In support of these predictions, a series of measurements has been
published the results of which are listed in the table \ref{tab_decay},
together with the half-life changes predicted by the Debye-H\"{u}ckel
model. %
\begin{table}
\caption{\label{tab_decay} Decay of radionuclides embedded in host metals}

\begin{ruledtabular}

\begin{tabular}{ccrcrr}
Ref.  & nuclide  & decay mode  & host  & prediction  & measurement\tabularnewline
\hline
\citep{limata06}  & $^{22}$Na  & 90\% $\beta^{+}$  & Pd  & 11\%  & ($1.2\pm0.2$)\%\tabularnewline
\citep{spillane07}  & $^{198}$Au  & 100\% $\beta^{-}$  & Au  & -34\%  & (-$4.0\pm0.7$)\%\tabularnewline
\citep{raiola07}  & $^{210}$Po  & 100\% $\alpha\,\,\,\,$  & Cu  & 3300\% & ($6.3\pm1.4$)\%\tabularnewline
\end{tabular}

\end{ruledtabular}
\end{table}
 The striking disagreement with the predictions have been attributed
to an oxygen layer build-up on the metal surface leading to an insufficient
implantation of the radioisotope %
\footnote{Publications concerning a change in $^{7}$Be lifetime are not taken
into account as $^{7}$Be decays via capture of s-wave electrons which
is not influenced by electron screening.%
}.

A recently published measurement \citep{ruprecht07,ruprecht08} where
$^{22}$Na was activated in Al (and therefore deeply implanted) clearly
shows a zero effect on a level of 0.04\%, again in striking disagreement
to the results by \citep{limata06} with a reported lifetime change
of $\left(1.2\pm0.2\right)\%$ (see table~\ref{tab_decay}). No description
has been given in \citep{limata06} how the data have been analyzed.
If the 511-keV annihilation line has been included in the analysis,
the results are certainly not correct (see \citep{canter74}). For
the $\alpha$ decay, even the observed 6\% \citep{raiola07} change
is surprising as embedding radioactive nuclei in metals and cooling
the samples to cryogenic temperatures is a routine procedure in low
temperature nuclear orientation (LTNO) experiments since several decades
\footnote{It should be noted that polonium is known to be very movable in metals
\citep{zastawny92}, therefore an alteration of the measured activity
could be due to changes in the polonium distribution.%
}. Stone \textit{et al.} \citep{stone06} studied in detail the expected
effect w.r.t.~$\alpha$ decay on complete decay chains starting with
$^{224}$Rn, $^{225}$Ra, and $^{227}$Ac and compared it with available
LTNO data. None of those data indicate any change of the lifetime
of any of the nuclei involved when they are implanted into Iron, neither
at room temperature nor when cooled to 20 millikelvin. The same applies
for $\beta$ active nuclei in multiple host metals, see \citep{stone06}
and references therein. The precision of these measurements is typically
1\% and less. Another follow-up measurement performed at ISOLDE/CERN
\citep{jeppesen07} focused on a possible change of the $^{221}$Fr
($\alpha$ decay) half-life when embedded in a metal and an insulator;
there is also no clear effect (50\% error) on a level of 0.3\%. Severijns
\textit{et al.} \citep{severijns07} investigated the $\alpha$ decay
of $^{253}$Es in Fe between 4~K and 50~mK and could not observe
any effect on a level of 2\%. Finally, also the $\beta^{-}$ decay
of $^{198}$Au embedded in Au and Al-Au has been measured independently
by three different groups \citep{goodwin08,kumar08,ruprecht08}, and
no lifetime change could be observed on a sub-percent level when the
sample was cooled to $\approx$10~K. The latest result \citep{ruprecht08}
was measured with a 30 times better error but the same conditions
as in \citep{spillane07}.

In conclusion, all the follow-up measurements are in agreement with
the theoretical expectations presented already in Sec.~\ref{sec:results.theo}.
The Debye-H\"{u}ckel screening can be applied only for temperatures
higher than the Fermi temperature being typically $10^{5}\,\kelvin$,
far above the evaporation temperature of metals. For lower temperatures
one should not observe any temperature dependence of the screening
energy.

However, another effect can be expected \citep{genf06}: a change
of the lifetime by just embedding the unstable nuclei into a metal,
but this requires an absolute measurement of the lifetime and therefore
much more experimental effort. A re-analysis of past lifetime measurements
data with respect to the chemical composition could also reveal such
a dependence. For instance, the lifetime of $^{238}$U has been determined
with electroplated samples (metallic uranium), with U$_{3}$O$_{8}$,
and other compounds, see \citep{schoen04,jaffey71}. Although the
measurements scatter by 1-2\%, no systematic enhancement of the decay
rate can be seen for the metallic uranium. As can be seen from the
aforementioned CERN measurement \citep{jeppesen07} the effect will
be small anyway even for high screening values but an evidence would
be a great contribution to a better understanding of the screening
mechanism from a very different approach.

\section{Conclusion}

We presented some new experimental electron screening energies for
d+d reactions taking place in different target material environments.
We applied a differential data analysis method which gains the maximum
information from the raw data. The method is independent of the unprecise
stopping power coefficients and the actual absolute value of the deuteron
number density in the targets. It enables the on-line monitoring of
the deuteron densities and the observation of short time deuteron
density profile changes. Thus, it allows for the recognition and rejection
of measurements with unwanted shifts in the density depth distribution
profile. Therefore, it adequately considers the special situation
of potentially highly mobile hydrogen in solid states where neither
a homogeneous nor a stable density distribution can be presupposed
any longer. The problem of the density dynamics is entangled with
the effects from the actual target composition, i.e.~the undesirable
density profile changes occur in targets with low hydrogen binding
ability, like many of the transition metals, at elevated temperatures
and heterogeneous targets with metal oxide or carbon layers or different
(relatively) thin metal layers. The formation of metal oxide layers
is inevitable in common high vacuum systems used in experimental nuclear
physics while the other unpropitious environments were produced deliberately.
Thorough investigation of the contamination layer formation showed
their momentousness and assured together with the differential analysis
method that our screening energy values ranging between $190-320\,\electronvolt$
represent lower limits. In addition the alteration of the inferred
screening energies due to layer formation under beam irradiation depends
on many parameters. Logically it makes no sense to measure larger
portions of the periodic table since any observed material dependence
results from differences in the chemical reactivity and related physico-chemical
properties for the contamination layer formation, unless this problem
is reliably solved. Different to the other two groups our high screening
energy results were achieved at high densities in the proximity of
the chemical stoichiometric ratio clearly without evidence for short
time density profile shifts. Whereas the high screening results of
the other groups were exclusively attained at low densities yielded
from the customary analysis of the total yields of the measurements
which is blind for the then happening density dynamics. The target
diagnosis methods are unusable because of their too bad resolution
and off-line application. So the inferred screening energies are conjecturally
simulated by the density dynamics. Utilization of explorative statistics
to the data sets including the temperature measurements sustains this
explanation while on the other hand the Debye-H\"{u}ckel hypothesis
is clearly falsified. It is likewise falsified from the theoretical
side since calculations performed within an improved dielectric function
theory predict only a weak material dependence of $U_{e}$ on the
valence electron density. The quantitative scale of the phenomenon
is not yet understood, since our analytical model still fails to describe
the values by at least a factor of 2. So further unidentified effects
play a role. Consequently any conclusion based on the alleged material
dependence of the inferred screening energies is premature. For it
the precise determination of the screening energies is demandable
which is only feasible in an ultra high vacuum system with pressures
well below $10^{-10}\,\hecto\pascal$, where only hydrogen and noble
gases are in the residual gas, and equipped with in-situ target diagnosis
techniques. We performed the first measurements under UHV conditions,
whose results confirm the previous measurements and the framework
of surface physics and chemical effects \citep{jphysg08}.

Nuclear reactions with heavier nuclei embedded in metallic environments
gave evidence for an alike enhanced screening effect. However, there
are analogue problems. The results for deuterated metals with $^{3}$He
projectiles are contradictory, most probably due to deuteron dynamics.
The data for Li nuclei are partially conflicting between the Tohoku
and the Bochum group, which used different target preparation techniques,
and can be attributed to inhomogeneous densities and inadequate diagnosis
techniques as well. The results, however, confirm theoretical predictions
based on the dielectric function theory concerning the free electron
density of the target material. The screening energy data for the
heavier nuclei V and Lu were obtained from a comparison between a
metal and a metal oxide powder target ignoring the hollow spaces in
the powder and its strong influences on beam stopping, thus disabling
conclusions.

As discussed, the predictions of the Debye-H\"{u}ckel hypothesis
given by the Bochum group for the temperature dependence of the radioactive
decay of embedded nuclei could not be verified by their own experiments;
the measured values are orders of magnitude below their predictions.
Moreover, their experimental results are in contradiction to all other
experiments, in particular the LTNO measurements of the past 30 years.
A material dependence is conceivable though a small effect. Otherwise
it would have been already discovered given that nuclei of importance
for nuclear technology have been investigated in multiple chemical
compounds including pure metals for decades.

\end{document}